\documentclass[acmsmall,screen]{acmart}

\usepackage{algpseudocode}
\usepackage{algorithm}
\usepackage{amsmath,amsfonts}
\usepackage{array}
\usepackage[caption=false,font=normalsize,labelfont=sf,textfont=sf]{subfig}
\usepackage{textcomp}
\usepackage{stfloats}
\usepackage{url}
\usepackage{verbatim}
\usepackage{graphicx}
\usepackage{zref}
\usepackage[normalem]{ulem}
\usepackage{tabularx}
\usepackage{setspace}
\usepackage{multirow}
\usepackage{booktabs}
\usepackage{listings}
\usepackage{xcolor}
\usepackage{makecell}
\usepackage{mdframed}
\usepackage{hyperref}
\usepackage[pagewise]{lineno}
\usepackage{booktabs} % 三线表宏包
\usepackage{amssymb}  % 提供 \checkmark
\usepackage{pifont}   % 提供 \ding{55} (叉号)

\lstdefinelanguage{Java}{
    keywords={public, private, new, null, true, false, throws, assert, return}
}
\lstdefinelanguage{JSON}{
    keywords={true, false, null, yes, no},
    sensitive=true,
    morecomment=[l]{\#},
    morestring=[b]{"},
    keywordstyle=\bfseries\color{blue},
    commentstyle=\color{gray}
}
\lstdefinelanguage{YAML}{
    keywords={true, false, null, yes, no},
    sensitive=true,
    morecomment=[l]{\#},
    morestring=[b]{"},
    keywordstyle=\bfseries\color{blue},
    commentstyle=\color{gray},
    stringstyle=\color{green}
}
\usepackage[justification=centering]{caption}
\usepackage{framed}
\usepackage{tcolorbox}
\tcbuselibrary{skins}

\newcounter{DaveCommentCounter}
\newcommand{\ddu}[1]{
    \stepcounter{DaveCommentCounter}
    \textcolor{blue}{\textit{/**Dave's comment [\arabic{DaveCommentCounter}]: I don't understand the intended meaning in the next sentence. Please revise/delete/explain. **/}}
}

\newcommand{\dns}[1]{
    \stepcounter{DaveCommentCounter}
    \textcolor{blue}{\textit{/**Dave's comment [\arabic{DaveCommentCounter}]: I'm not sure that I have captured the intended meaning in the next sentence. Please check/confirm. **/}}
}

\newcounter{RubingCommentCounter}
\newcounter{SidaDengCounter}
\AtBeginDocument{%
  }

\setcopyright{acmlicensed}
\copyrightyear{2026}
\acmYear{2026}
\acmDOI{XXXXXXX.XXXXXXX}

\acmJournal{JACM}
\acmVolume{88}
\acmNumber{8}
\acmArticle{888}
\acmMonth{6}
\acmISBN{978-1-4503-XXXX-X/18/06}

\begin{document}

\title{MASTOR: A Multi-Agent Approach to Semantic Test Oracle Generation for RESTful APIs}

\author{Sida Deng}
\email{3220002511@student.must.edu.mo}
\orcid{0009-0006-2819-3862}
\affiliation{
  \institution{School of Computer Science and Engineering, Macau University of Science and Technology}
  \city{Macao SAR}
  \country{China}
  \postcode{999078}
}

\author{Rubing Huang}
\authornotemark[1]
\email{rbhuang@must.edu.mo}
\orcid{0000-0002-1769-6126}
\affiliation{
  \institution{School of Computer Science and Engineering, Macau University of Science and Technology}
  \city{Macao SAR}
  \country{China}
  \postcode{999078}
}
\affiliation{
  \institution{Macau University of Science and Technology Zhuhai MUST Science and Technology Research Institute}
  \city{Zhuhai}
  \country{China}
  \postcode{519099}
}

\author{Zhenzhen Yang}
\email{3240002362@student.must.edu.mo}
\orcid{0009-0009-9358-7675}
\affiliation{
  \institution{School of Computer Science and Engineering, Macau University of Science and Technology}
  \city{Macao SAR}
  \postcode{999078}
  \country{China}  
}

\author{Man Zhang}
\authornote{Corresponding author.}
\email{manzhang@buaa.edu.cn}
\orcid{0000-0003-1204-9322}
\affiliation{
  \institution{State Key Laboratory of Complex \& Critical Software Environment, Beihang University}
  \city{Beijing}
  \country{China}
  \postcode{100191}
}

% \author{Dave Towey}
% \email{dave.towey@nottingham.edu.cn}
% \orcid{0000-0003-0877-4353}
% \affiliation{
%   \institution{School of Computer Science, University of Nottingham Ningbo China}
%   \city{Ningbo}
%   \country{China}
%   \postcode{999078}
% }

\author{Xuan Xie}
\email{xiexuan@must.edu.mo}
\orcid{0000-0003-3981-8515}
\affiliation{
  \institution{School of Computer Science and Engineering, Macau University of Science and Technology}
  \city{Macao SAR}
  \country{China}
  \postcode{999078}
}

\author{Rongcun Wang}
\email{rcwang@cumt.edu.cn}
\orcid{0000-0002-9685-9893}
\affiliation{
  \institution{School of Computer Science and Technology, China University of Mining and Technology}
  \city{Xuzhou}
  \country{China}
  \postcode{221116}
}

\renewcommand{\shortauthors}{Deng et al.}

\newcommand{\placeholder}[1]{\textcolor{blue}{#1}}

\begin{abstract}
Test oracle generation is receiving increasing attention in RESTful API testing.
Existing automated approaches commonly rely on simple checks, such as HTTP status codes, runtime failures, and schema conformance.
However, these checks are insufficient for detecting semantic faults, business logic violations, and state-dependent behavioral inconsistencies encoded in API implementations.
Motivated by such a fact, we propose MASTOR, a \underline{\textbf{\textit{M}}}ulti-\underline{\textbf{\textit{A}}}gent approach for generating \underline{\textbf{\textit{S}}}emantic \underline{\textbf{\textit{T}}}est \underline{\textbf{\textit{O}}}racles for \underline{\textbf{\textit{R}}}ESTful APIs, based on the implementation source code.
MASTOR consists of the following two phases: \textit{source analysis} and \textit{oracle generation}.
More specifically, the former phase employs a source extraction agent to construct a source context for each endpoint operation by analyzing a transitive import closure of relevant source files; however, the latter phase employs two parallel oracle-generation paths over the collected source contexts: 
a single-operation path that produces status and field oracles per endpoint operation, while a multi-operation path that generates behavioral consistency oracles for operation sequences by leveraging cross-operation semantic associations identified during the source analysis phase.
Both paths apply a challenger-agent review, in which a dedicated reviewer identifies weaknesses in the generated oracles and issues improvement hints to guide targeted regeneration, followed by oracle normalization to filter out structurally invalid oracles before contributing to the oracle suite.
To evaluate the performance of MASTOR, we conduct a series of experiments on a benchmark comprising 13 open-source RESTful API projects with 296 endpoint operations and 251,303 lines of code, drawn from the WFD (Web Fuzzing Dataset) and PRAB (Public RESTful API Benchmark) datasets.
MASTOR achieves an average mutation score of 75.4\% across all 13 APIs, generating 10,022 oracles.
The generated oracles are translated into executable assertions via \texttt{ToJUnit} and \texttt{ToPostmanAssertify}, and into human-readable descriptions via \texttt{ToReadable} for manual inspection by testers.
In a baseline comparison on 50 selected endpoint operations using status and field oracle types only, MASTOR outperforms Direct Prompting by 30.1 percentage points (69.9\% vs.\ 39.8\%) and SATORI by 49.4 percentage points (69.9\% vs.\ 20.5\%).
\end{abstract}




\maketitle

\section{Introduction
\label{sec:1}}
RESTful APIs have become the dominant architectural style for building modern web services and distributed applications. 
As service-oriented systems continue to grow in scale and complexity, ensuring the correctness and reliability of RESTful APIs has become increasingly important. 
Automated testing techniques have therefore received substantial attention in both research and practice~\cite{golmohammadiTestingRESTfulAPIs2023b,zhangOpenProblemsFuzzing2023a,garciaAdvancesWebAPI2023}.
Recent work in RESTful API testing has primarily focused on automated test case generation, including fuzzing-based~\cite{atlidakisRESTlerAutomaticIntelligenta,
diasFuzzTheRESTIntelligentAutomated2024,zhengRESTLessEnhancingStateoftheArt2024,rooijakkersWuppieFuzzCoverageGuidedStateful2026,fernandesCaseStudyApplying2025}, search-based~\cite{zhangAdaptiveHypermutationSearchBased2021,arcuriAdvancedWhiteBoxHeuristics2024,golmohammadiEnhancingWhiteBoxSearchBased2023,stallenbergImprovingTestCase2021a,ghianniSearchBasedFuzzingRESTful2025}, and approaches based on Large Language Models (LLMs)~\cite{kimMultiAgentApproachREST2025a,stennettAutoRestTestToolAutomated2025b,kimLlamaRestTestEffectiveREST2025a,barradasCombiningTSLLLM2025,nooyensTestAmplificationREST2025,besjesAgenticLLMsREST2025}.
Most of these techniques operate on a machine-readable interface contract, typically an OpenAPI Specification (OAS), which describes the available endpoint operations, their parameters, and their response schemas.
A separate line of work generates the OAS itself directly from API source code, using either static analysis~\cite{huangGeneratingRESTAPI2024,lercherGeneratingAccurateOpenAPI2024a} or LLMs~\cite{dengLRASGenLLMbasedRESTful2026}, recovering endpoint entities that developer-provided specifications miss.
% \rbc{Here, for each item, there are so many references, try to use the latest one or two references.}
While these techniques improve input diversity and coverage, their verification mechanisms are often limited to simple checks, such as HTTP status codes, runtime failures, and schema conformance.
Although such checks are lightweight and easy to automate, they are insufficient for detecting semantic faults, business logic violations, and state-dependent behavioral inconsistencies embedded in API implementations.
In practice, many RESTful API defects do not manifest as crashes or invalid schemas, but rather as incorrect response status codes across different execution paths, malformed or missing response body fields, and data inconsistencies between related endpoint operations.

Test oracle generation is therefore emerging as a critical research direction for RESTful API testing~\cite{alonsoAGORAAutomatedGeneration2023,alonsoTestOracleGeneration2025,alonsoSATORIStaticTest2025a}.
However, constructing semantic oracles for RESTful APIs is inherently challenging because correctness spans both per-endpoint behavior and cross-endpoint relationships.
Per-endpoint correctness involves status codes that vary by execution path and field-level constraints on response body properties.
Cross-endpoint correctness further involves behavioral consistency among operations that share data dependencies, such as identifier propagation through path parameters and consistent resource representations across related endpoints.
These correctness dimensions are distributed across multiple source code components (controller methods, service-layer logic, repository interfaces, and data model definitions), making it difficult for a single analytical pass to extract all relevant constraints from source evidence alone.
Each layer requires a different analysis context: extracting parameter constraints from a controller handler demands different input than inferring cross-operation data flow from a set of resource endpoints.
A monolithic LLM query cannot simultaneously maintain focused, layer-specific context across all layers without incurring hallucinations due to context overload.
This motivates a multi-agent approach that decomposes the analysis into specialized subtasks, each operating on a precisely bounded context, and coordinates them over a shared structured representation of the source code.

In this paper, we propose MASTOR, a \underline{\textbf{\textit{M}}}ulti-\underline{\textbf{\textit{A}}}gent approach for generating \underline{\textbf{\textit{S}}}emantic \underline{\textbf{\textit{T}}}est \underline{\textbf{\textit{O}}}racles for \underline{\textbf{\textit{R}}}ESTful APIs, based on the implementation source code.
Unlike approaches~\cite{alonsoSATORIStaticTest2025a, atlidakisRESTlerAutomaticIntelligenta}
that rely solely on interface-level verification, MASTOR analyzes complete RESTful API projects, including controller handlers, service-layer logic, and data model definitions.
MASTOR operates in two phases: Source Analysis and Oracle Generation.
In the Source Analysis phase, a Source Extraction Agent analyzes each endpoint operation by reading a transitive import closure of relevant source files, leveraging LLM capabilities to interpret framework annotations, inheritance, and generic types.
The output is a Source context that captures the input parameter constraints and response schema for each operation.
In the Oracle Generation phase, two parallel paths operate over the collected Source contexts.
The single-operation path generates two categories of oracles per endpoint, namely status oracles (HTTP response codes per execution path) and field oracles (response body field constraints), using four complementary strategies that cover valid and invalid inputs from both forward and backward reasoning directions (Section~\ref{sec:4.4}).
The multi-operation path generates behavioral consistency oracles for operation sequences from the cross-operation semantic associations identified during Source Analysis.
Both paths apply a challenger-agent review, in which a dedicated reviewer identifies weaknesses in the generated oracles and issues improvement hints to guide targeted regeneration, followed by oracle normalization to filter out structurally invalid oracles before contributing to the oracle suite.
Every generated assertion is grounded in a source code element identified during extraction.

We evaluate MASTOR on a benchmark of 13 RESTful API projects drawn from the WFD (Web Fuzzing Dataset)~\cite{sahinWFCWFDWeb2025} and PRAB (a Public RESTful API Benchmark)~\cite{decropPublicBenchmarkREST2025a} datasets.
To construct the benchmark, we surveyed the SUT projects used across more than 60 RESTful API testing studies, ranked them by frequency of appearance, and retained all 13 projects referenced by at least eight studies.
The full selection criteria and candidate list are described in Section~\ref{sec:5}.
The evaluation covers 296 endpoint operations across 251,303 lines of code.
An oracle generation experiment (RQ1) across all 13 subject APIs shows that MASTOR achieves an average mutation score of 75.4\%, with individual API scores ranging from 69.0\% to 95.9\%, generating 10,022 oracles.
The generated oracles are translated into executable assertions via \texttt{ToJUnit} and \texttt{ToPostmanAssertify}, and into human-readable descriptions via \texttt{ToReadable} for manual inspection by testers.
A baseline comparison (RQ2) on 50 selected endpoint operations using status and field oracle types only shows that MASTOR outperforms Direct Prompting by 30.1 percentage points (pp) (69.9\% vs.\ 39.8\%) and SATORI by 49.4 pp (69.9\% vs.\ 20.5\%).
An ablation study (RQ3) confirms that both architectural components (Multi-Op Oracle Generation and ChallengerAgent) contribute positively, with Multi-Op Oracle Generation providing the larger contribution.
An efficiency analysis (RQ4) across all 13 APIs shows that the median inference cost is \$0.56 per API using DeepSeek V4 Pro and Qwen3.6-Plus, and that computational cost scales with endpoint count and the size of each operation's transitive import closure.

The contributions of this paper are summarized as follows:
\begin{itemize}
    \item [(1)] We present MASTOR, a multi-agent approach to generating semantic test oracles for RESTful APIs directly from implementation source code.

%\item [2)] We design a layered, evidence-driven collaboration architecture consisting of an LLM-assisted analysis agent for inter-procedural slicing, a deterministic static evidence builder, and four oracle generation agents that generate and verify semantic assertions grounded in explicit source code elements.
    \item [(2)] We design a two-phase architecture in which the Source Analysis phase extracts per-operation Source contexts, and the Oracle Generation phase runs parallel single-operation and multi-operation paths, each with ChallengerAgent review and Oracle Normalization.

%\item [3)] We develop a systematic integration strategy that incorporates existing invariant definitions (e.g., AGORA) into a source-code-evidence-guided oracle generation workflow, enabling invariant instantiation and pruning based on concrete implementation source code.
    \item [(3)] We introduce Source context as a structured intermediate representation of per-operation source facts, and the oracle suite as the final merged oracle output of all generated oracles.

    \item [(4)] We implement three converters (\texttt{ToJUnit}, \texttt{ToPostmanAssertify}, and \texttt{ToReadable}), enabling automated test execution, CI/CD integration, and manual oracle inspection.

    \item [(5)] We conduct a series of experiments on 13 RESTful API projects, demonstrating that MASTOR achieves a strong mutation score and outperforms both Direct Prompting and SATORI by substantial margins.
    Ablation experiments confirm the contribution of each architectural component, and efficiency analysis shows that inference cost remains practical at scale.

    \item [(6)] We release MASTOR as an open-source artifact, including the full implementation, benchmark datasets, and the complete oracle suite generated for all 13 subject APIs~\cite{mastorGitHub}.
    The oracle suite constitutes a reusable semantic oracle collection for the benchmark APIs, enabling independent replication and serving as a baseline for future oracle generation research.

\end{itemize}

The remainder of this paper is organized as follows.
Section~\ref{sec:background} presents background and motivation.
Section~\ref{sec:3} presents a motivating example illustrating the semantic oracle problem.
Section~\ref{sec:method} describes the design and implementation of MASTOR.
Section~\ref{sec:5} describes the experimental methodology.
Section~\ref{sec:evaluation} reports the experimental results.
Section~\ref{sec:discussion} discusses the findings, applicability, and threats to validity.
Section~\ref{sec:related} reviews related work, and Section~\ref{sec:conclusion} concludes the paper and outlines future work.

\section{Background
\label{sec:2}\label{sec:background}}
In this section, we provide background information on RESTful API testing, the test oracle problem, and multi-agent decomposition.

\subsection{RESTful API Testing and Verification Limitations}

The resource-oriented design, stateless interaction model, and standardized HTTP semantics of RESTful APIs support scalable service composition, but also introduce testing challenges related to state transitions, data persistence, and cross-layer logic consistency~\cite{fieldingArchitecturalStylesDesign2000}.

Automated RESTful API testing has received substantial research attention. Existing approaches primarily focus on test case generation, including fuzzing-based strategies~\cite{atlidakisRESTlerAutomaticIntelligenta,godefroidIntelligentRESTAPI2020a}, search-based exploration~\cite{zhangAdaptiveHypermutationSearchBased2021,arcuriAdvancedWhiteBoxHeuristics2024}, model-based generation~\cite{liuMorestModelbasedRESTful2022a}, and LLM-assisted techniques~\cite{kimLeveragingLargeLanguage2024a,kimMultiAgentApproachREST2025a,hanMASTESTLLMBasedMultiAgent2025}. These methods aim to improve input diversity, structural coverage, and endpoint reachability.

However, verification mechanisms in many existing tools remain relatively lightweight. Commonly adopted checks include HTTP status code validation, runtime crash detection, and response-schema conformance~\cite{golmohammadiTestingRESTfulAPIs2023b,garciaAdvancesWebAPI2023}. While such checks are easy to automate and scale well in continuous testing environments, they do not capture semantic violations embedded in the implementation. In practice, RESTful API faults frequently manifest as incorrect state transitions, unintended persistence effects, inconsistent field updates, or violations of implicit business constraints rather than explicit failures observable at the interface level.
The limitations of interface-level verification highlight the need for semantic test oracles that reason about expected behavior beyond superficial response properties.
Constructing such oracles requires analyzing implementation source code to extract behavioral properties encoded in input constraints, branching conditions, and exception handling rather than in API specifications alone.
% \wrc{This paragraph is too short. It is recommended to merge it with the previous paragraph.}
% \dsd{merged}

\subsection{Test Oracle Problem and Semantic Constraints}

A test oracle determines whether the observed behavior of a system under test is correct with respect to expected behavior~\cite{barrOracleProblemSoftware2015}. Constructing automated oracles is widely recognized as one of the central challenges in software testing research. 

Existing oracle construction strategies can be broadly categorized into specification-based, model-based, differential, invariant-based, and implementation-based approaches~\cite{barrOracleProblemSoftware2015}. Specification-based oracles rely on formal contracts or API documentation, but such documents are often incomplete or imprecise in practice. Model-based techniques require manually constructed behavioral models, which are costly to maintain for evolving service-oriented systems. Differential approaches depend on reference implementations or cross-version comparison, which are not always available.

Invariant-based methods infer properties that should hold across executions, such as value ranges, structural relationships, or data consistency constraints~\cite{ernstDaikonSystemDynamic2007}. Systems such as AGORA~\cite{alonsoAGORAAutomatedGeneration2023}, an invariant-template-based oracle generation tool for object-oriented programs, demonstrate that invariant templates can effectively capture common implementation-level constraints. Nevertheless, invariant inference alone does not guarantee semantic completeness for RESTful APIs, as endpoint correctness may depend on control-flow conditions, exception handling logic, inter-layer interactions, and persistence-side effects that extend beyond isolated value properties.
Specification-based approaches that leverage LLMs have emerged more recently.
SATORI~\cite{alonsoSATORIStaticTest2025a} is a representative example: it is a black-box approach that uses an LLM to analyze the OpenAPI Specification~\cite{openapispecificationOpenAPISpecification} of a REST API, inferring expected response field behavior from field names and descriptions, and generating status and field oracles without accessing the implementation source code.
Implementation-based approaches
 derive oracle constraints directly from the program's source code, reading control-flow predicates, parameter validation logic, and exception-handling paths to ground assertions in the actual behavior of the implementation rather than in a separately maintained specification or model.

For RESTful APIs, endpoint behavior is typically distributed across controller handlers, service-layer components, repository interfaces, and serialization mechanisms. Semantic correctness, therefore, depends on the combined effects of input validation, branching conditions, state mutation, exception propagation, and response construction. Many of these properties are encoded implicitly in the source code rather than explicitly declared in interface specifications.

These characteristics suggest that effective oracle generation for RESTful APIs should (1) analyze complete implementation source code rather than API specification alone, (2) extract structured evidence capturing the constraints encoded in implementation source files, and (3) ensure that synthesized assertions remain grounded in concrete program elements.
MASTOR adopts the implementation-based approach and satisfies all three requirements: it analyzes complete Java RESTful API projects via transitive import closure, extracts per-operation structured representations of input constraints and response schemas, and grounds every generated assertion in a source code element identified during extraction.

\subsection{Multi-Agent Decomposition for Structured Code Reasoning}

Recent advances in collaborative agent architectures suggest that complex reasoning tasks can benefit from role specialization and decomposition~\cite{kimMultiAgentApproachREST2025a,hanMASTESTLLMBasedMultiAgent2025}. Instead of relying on monolithic reasoning, multi-agent approaches divide tasks into coordinated subtasks handled by specialized components, enabling structured information exchange and iterative refinement.

Oracle generation from RESTful API source code naturally involves multiple reasoning dimensions, including status-code reasoning across execution paths, field-level constraint inference on response bodies, and cross-operation consistency verification. These subtasks differ in required context granularity and reasoning strategy. Decomposing oracle generation into coordinated agent roles can improve modularity, interpretability, and controllability compared to single-pass reasoning approaches.

In the context of source-code-grounded oracle generation, a multi-agent architecture also facilitates explicit evidence tracking. By associating intermediate reasoning steps with concrete program source code, the system can reduce unsupported assertions and improve logical consistency. This design perspective motivates the collaborative structure adopted in MASTOR, which pairs specialized oracle-generation agents with dedicated reviewer agents to ensure that every assertion is grounded in explicit source code evidence.
MASTOR instantiates this architecture in two phases: a Source Analysis phase that constructs a structured intermediate representation of each endpoint's input constraints and response schema from the implementation source files, and an Oracle Generation phase that coordinates specialist agents to generate oracles and reviewer agents to verify them, both operating over that shared structured representation.

\section{A Motivating Example
\label{sec:3}}
We use two endpoint operations from the \texttt{restcountries} project in the WFD dataset~\cite{sahinWFCWFDWeb2025} to illustrate one instance of each oracle category.
Listing~\ref{lst:motivating-countries} indicates the \texttt{GET /v2/alpha/\{alphacode\}} endpoint operation and its field-filtering helpers; while Listing~\ref{lst:motivating-stripe} indicates the \texttt{POST /contribute} endpoint operation from the same project.

\begin{lstlisting}[
language=Java,
caption={``\texttt{GET /v2/alpha/\{alphacode\}}'' endpoint operation and field-filtering helpers from \texttt{CountryRestV2.java} in \texttt{restcountries} API.},
label={lst:motivating-countries}]
@GET
@Path("alpha/{alphacode}")
public Object getByAlpha(@PathParam("alphacode") String alpha,
                         @QueryParam("fields") String fields) {
    if (isEmpty(alpha) || alpha.length() < 2 || alpha.length() > 3) {
        return getResponse(Response.Status.BAD_REQUEST);
    }
    Country country = CountryService.getInstance().getByAlpha(alpha);
    if (country != null) { return parsedCountry(country, fields); }
    return getResponse(Response.Status.NOT_FOUND);
}
private Object parsedCountry(Country country, String fields) {
    if (fields == null || fields.isEmpty()) { return country; }
    else { return getCountryJson(country,
        Arrays.asList(fields.split(ICountryRestSymbols.SEMICOLON))); }
}
private String getCountryJson(Country country, List<String> fields) {
    JsonObject jsonObject = parser.parse(gson.toJson(country)).getAsJsonObject();
    for (String field : getExcludedFields(fields)) { jsonObject.remove(field); }
    return jsonObject.toString();
}
private List<String> getExcludedFields(List<String> fields) {
    List<String> excluded = new ArrayList<>(Arrays.asList(COUNTRY_FIELDS));
    excluded.removeAll(fields);
    return excluded;
}
private static final String[] COUNTRY_FIELDS = {
    "name", "alpha2Code", "alpha3Code", "capital",
    "currencies", "languages", /* 18 more fields */ };
\end{lstlisting}

\begin{lstlisting}[float=t,
language=Java,
caption={``\texttt{POST /contribute}'' endpoint operation from \texttt{StripeRest.java} in \texttt{restcountries} API.},
label={lst:motivating-stripe}]
@POST
public Object contribute(Contribution contribution) {
    if (contribution == null || isBlank(contribution.getToken())) {
        return getResponse(Response.Status.BAD_REQUEST);
    }
    Map<String, Object> params = new HashMap<>();
    params.put("amount", contribution.getAmount());
    params.put("currency", "eur");
    params.put("description", "REST Countries");
    params.put("source", contribution.getToken());
    try {
        Charge.create(params);
    } catch (AuthenticationException | InvalidRequestException
           | CardException | APIConnectionException | APIException e) {
        return getResponse(Response.Status.BAD_REQUEST);
    }
    return getResponse(Response.Status.ACCEPTED);
}
\end{lstlisting}
% \wrc{Here, I have adjusted the position of the following listings. Replacing float=t with float=!htbp.}

At the interface level, a correct \texttt{GET /v2/alpha/\{alphacode\}} request is expected to return HTTP status code \texttt{200} with a JSON object representing the matched country.
For \texttt{POST /contribute}, a successful request yields HTTP status code \texttt{202} with a fixed response body.
Conventional automated tests would verify these response codes and schema conformance, deeming execution correct if these observable properties hold.

However, the semantic correctness of each endpoint operation depends on implementation-level constraints that are not declared in the API specification.
First, the \texttt{GET /v2/alpha/\{alphacode\}} operation enforces input format constraints encoded in the source code.
If the alpha code is absent, shorter than two characters, or longer than three, the operation returns HTTP status code \texttt{400}; if the code is syntactically valid but matches no country, it returns HTTP status code \texttt{404}.
These conditions are represented as control-flow predicates in the implementation.
A correct status oracle relates input characteristics to expected response codes across all execution paths.

Second, the \texttt{fields} query parameter determines which fields appear in the response body.
When \texttt{fields} is specified, \texttt{parsedCountry} delegates to \texttt{getCountryJson}, which calls \texttt{getExcludedFields} to compute a removal list: starting from the full set of 24 \texttt{COUNTRY\_FIELDS}, it removes the requested fields and deletes the remainder from the serialized JSON object.
The resulting response should contain exactly the fields named in the query parameter.
An incorrect implementation (for example, one that inverts the inclusion logic or applies the filter to the wrong set) would still return HTTP status code \texttt{200} with syntactically valid JSON, passing schema validation while violating the intended field contract.
A correct field oracle asserts the exact composition of the response body as a function of the \texttt{fields} input.

Third, the \texttt{POST /contribute} endpoint operation performs a side effect whose correctness cannot be inferred from the HTTP response.
A successful invocation returns HTTP status code \texttt{202} with a body containing only a status code and reason phrase.
The actual operation (a Stripe charge submitted with the caller-provided \texttt{amount}, a hardcoded \texttt{currency} of \texttt{"eur"}, and a fixed \texttt{description}) is absent from the response.
An incorrect implementation could substitute the provided amount, use a different currency, or omit the description, while still returning HTTP status code \texttt{202} and passing schema validation.
Detecting such a defect lies beyond the reach of single-operation response inspection: the correctness of the submitted charge parameters can only be verified by comparing them against the values actually received by the external service or by checking a subsequent retrieval operation. This example illustrates a general class of semantic faults affecting operations whose primary side effects are invisible in their HTTP responses.

These examples demonstrate that semantic properties of RESTful endpoints span three reasoning dimensions: input-driven branching conditions that determine status codes, input-dependent constraints on the composition of response fields, and operation-sequence requirements for verifying side effects whose correctness is not observable in a single response.
All three are encoded in the control flow and call structure of the source code rather than in API specifications alone.

Detecting violations requires synthesizing test oracles grounded in the implementation source code.
An effective oracle connects control-flow predicates (e.g., length guards), field-set computations (e.g., inclusion logic), and external call arguments (e.g., charge parameters) to observable assertions about expected outcomes.
These challenges motivate the design of MASTOR, which builds a Source context from each operation's implementation source bundle and coordinates specialized agents to synthesize evidence-grounded oracles covering all three reasoning dimensions.

\section{Approach
\label{sec:4}\label{sec:method}}
In this section, we mainly provide our proposed approach, MASTOR, including its architecture overview and core components.

\subsection{Overview
\label{sec:4.1}}

MASTOR is a multi-agent approach to automated RESTful API oracle generation.
Its core design principle is \textit{decomposition amplifies reasoning}: MASTOR decomposes the task across specialized agents, each with a narrowly scoped responsibility and precisely bounded context, rather than issuing a single unstructured LLM query.
Rule-based components handle tasks with fixed structure and no semantic ambiguity; LLMs handle tasks where correctness requires semantic interpretation.
Section~\ref{sec:4.2} details agent roles and this division of labor.

MASTOR adopts a blackboard-based multi-agent architecture~\cite{hayesrothBlackboardArchitectureControl1985}.
Agents coordinate exclusively through a shared OutputStore, a thread-safe blackboard holding Source contexts and generated oracles indexed by operation identifier, and submit tasks to a MessageBus (bounded thread pool) without calling each other directly.
This architecture differs from a linear pipeline in three respects.
First, agent instances run independently, so one operation's failure does not halt others.
Second, within each phase, agents process all endpoint operations or operation groups in parallel rather than sequentially.
Third, each oracle-generation agent is paired with a dedicated ChallengerAgent that performs one review pass, emitting improvement hints that guide a single targeted regeneration; this review-and-regeneration step is absent from any linear pipeline.

\begin{figure}[!b]
  \centering
  \includegraphics[width=\textwidth]{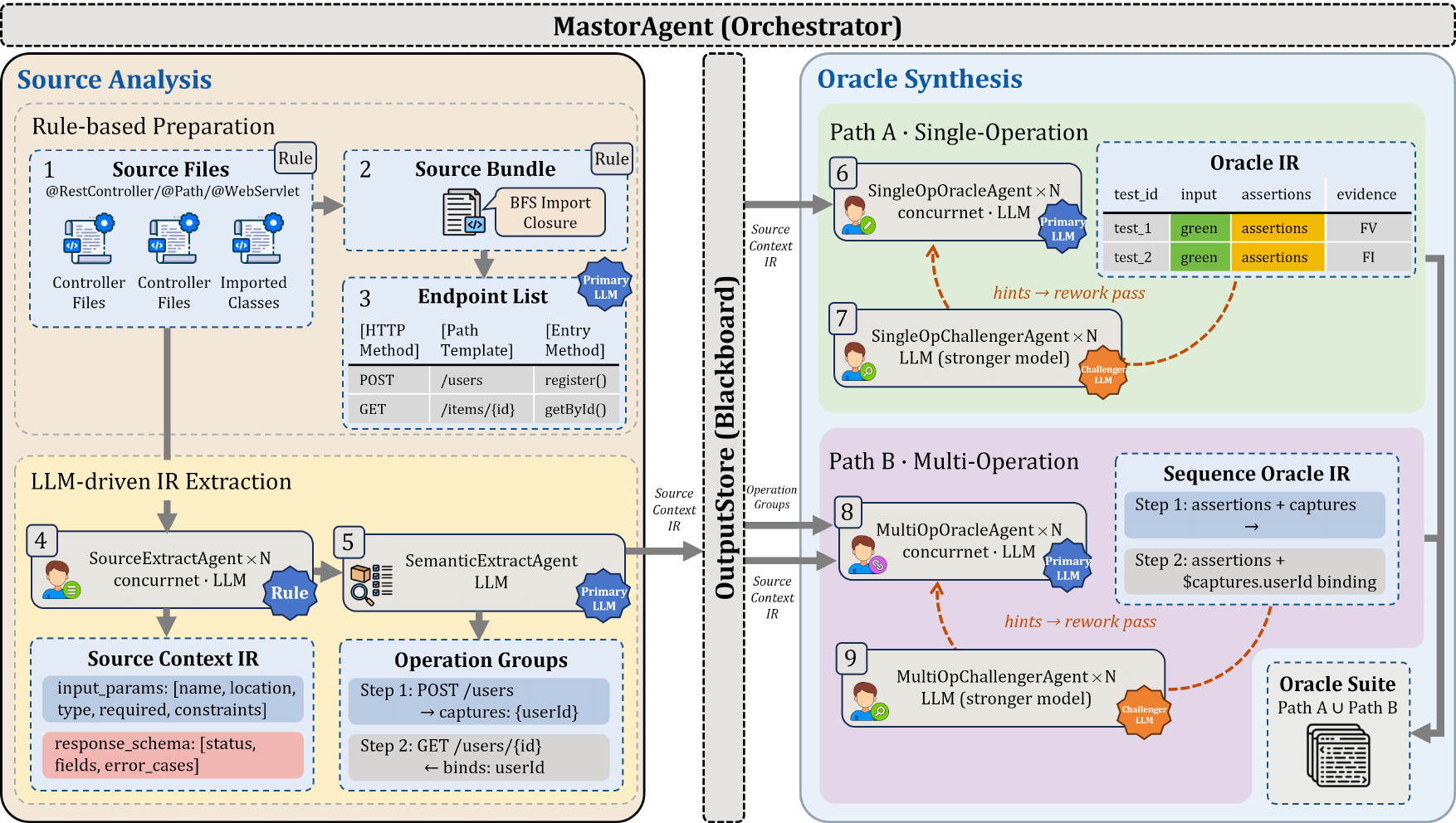}
  \caption{Overview of MASTOR. Within each phase, all agent tasks execute in parallel (concurrent arrows not shown for clarity).
  }
  \Description{A pipeline diagram of the MASTOR approach showing three phases: Phase A (Source Analysis) with an Endpoint Discovery agent and Source Extraction agents; Phase B (Oracle Generation) with Single-Op and Multi-Op oracle generation agents and a Challenger review agent; and Phase B output feeding into an oracle suite. Arrows indicate data flow between agents within and across phases.}
  \label{fig:overview}
\end{figure}

Figure~\ref{fig:overview} illustrates the workflow overview of the MASTOR approach.
Source code enters from the left: the project source tree is scanned, entry files are identified, and a transitive import closure is built for each endpoint operation to form source bundles.
In the Source Analysis phase, \texttt{SourceExtractionAgent} processes these bundles to produce one Source context per operation, while \texttt{SemanticExtractAgent} identifies cross-operation associations.
The Oracle Generation phase then runs two concurrent paths.
The single-op path generates per-operation oracles using four complementary strategies, followed by \texttt{SingleOpChallengerAgent} review and deterministic normalization.
The multi-op path consumes the operation groups from \texttt{SemanticExtractAgent}, generates multi-operation oracles, and applies the same review-and-normalize process.
Both paths write their oracles to the shared OutputStore; after both paths complete, the oracles are merged and exported as the oracle suite.

\subsection{Agents
\label{sec:4.2}}

Following the convention of LLM-based multi-agent frameworks~\cite{wuAutoGenEnablingNext2023,yaoReActSynergizingReasoning2023}, we use the term \textit{agent} to denote a self-contained execution unit with a specialized role, a bounded LLM context, and a defined read/write contract with the shared blackboard.
This usage does not imply autonomous goal-seeking; agents are coordinated by a deterministic orchestrator and communicate exclusively through the OutputStore.

Two structured representations govern information flow across agents.
The Source context encodes the input parameters and response schema for a single endpoint operation; it is written by \texttt{SourceExtractionAgent} and read (never modified) by all downstream agents.
To separate certainty from conjecture, the Source context carries two levels of fields: \texttt{source-verified} fields populated from concrete annotations and method bodies, and \texttt{pending} fields that hold OAS-declared items the source code does not substantiate (for example, error status codes returned only by a framework-level exception mapper).
Oracle-generation agents read only the \texttt{source-verified} fields, ensuring every generated assertion traces to a concrete source code element.
The mechanism by which \texttt{SourceExtractionAgent} populates these two levels and the precision-biased design rationale are detailed in Section~\ref{sec:4.3}.
Each generated oracle encodes test inputs, assertions, and source evidence, written by oracle-generation agents and reviewed by challenger agents.
After normalization and review, all oracles from both paths are merged into the oracle suite. This final output is then translated into executable test code and human-readable descriptions by the three converters (available at~\cite{mastorGitHub}).

\begin{table}[!b]
\centering
\footnotesize
\caption{MASTOR Agent Roles and Responsibilities}
\label{tab:agent_roles}
\setlength{\tabcolsep}{2mm}
\begin{tabular}{llp{8cm}}
\hline
\textbf{Role} & \textbf{Agent} & \textbf{Responsibility} \\
\hline
\multirow{4}{*}{Specialist}   & \texttt{SourceExtractionAgent}   & Extract input parameter constraints and response schema from source bundle \\
                              & \texttt{SemanticExtractAgent}    & Detect cross-operation semantic associations; produce ordered operation groups \\
                              & \texttt{SingleOpOracleAgent}     & Generate per-operation test oracles using four-strategy coverage \\
                              & \texttt{MultiOpOracleAgent}      & Generate multi-operation oracles for one operation group \\
\hline
\multirow{2}{*}{Reviewer}     & \texttt{SingleOpChallengerAgent} & Review single-op oracle suite; emit actionable improvement hints \\
                              & \texttt{MultiOpChallengerAgent}  & Review multi-operation oracle suite; emit actionable improvement hints \\
\hline
Orchestrator                  & \texttt{MastorAgent}             & Schedule agents, monitor OutputStore, route phase transitions \\
\hline
\end{tabular}
\end{table}

MASTOR's seven agents are organized into three roles: \textit{Specialist} agents extract source-code information and synthesize test oracles; \textit{Reviewer} agents provide challenger review; the \textit{Orchestrator} schedules agents and coordinates phase transitions without performing semantic inference.
Table~\ref{tab:agent_roles} lists each agent with its role and primary responsibility.
\begin{itemize}
    \item \textbf{Specialist agents} perform the substantive extraction and generation work. More specifically, \texttt{SourceExtractionAgent} and \texttt{SemanticExtractAgent} operate in the Source Analysis phase, transforming raw source files into structured intermediate representations consumed by all downstream agents (Section~\ref{sec:4.3}). \texttt{SingleOpOracleAgent} and \texttt{MultiOpOracleAgent} operate in the Oracle Generation phase, generating test oracle suites from the extracted Source context along single-operation and multi-operation coverage dimensions respectively (Section~\ref{sec:4.4}).

    \item \textbf{Reviewer agents} provide challenger review. Each is paired with a specialist: \texttt{SingleOpChall-\\engerAgent} reviews single-operation oracle suites; \texttt{MultiOpChallengerAgent} reviews multi-operation oracle suites. Both emit natural-language improvement hints without producing corrected oracles; the paired specialist performs one targeted regeneration pass guided by those hints, revising only the flagged oracles and preserving all others.

    \item \textbf{Orchestrator.} \texttt{MastorAgent} coordinates agent execution without performing semantic inference (\hyperref[OG5]{\textit{Scheduling}}). It submits tasks to the MessageBus, monitors the OutputStore to determine when each phase is complete, and isolates agent failures so that one operation's failure does not halt others. Sequencing between phases is deterministic; concurrency within each phase is managed by the MessageBus.

\end{itemize}

\subsubsection{LLM and Rule-based Division of Labor}
Table~\ref{tab:llm-algo-boundary} shows how each subtask within these roles is assigned to either an LLM or a rule-based component.

\begin{table}[!b]
\centering
\footnotesize
\caption{LLM-Rule-Based Division of Labor Across MASTOR Components}
\label{tab:llm-algo-boundary}
\setlength{\tabcolsep}{1.5mm}
\begin{tabular}{p{3cm}p{3.5cm}p{2cm}p{4cm}}
\hline
\textbf{Agent} & \textbf{Subtask} & \textbf{Executor} & \textbf{Reason} \\
\hline
\multicolumn{4}{l}{\textit{Source Analysis}} \\[2pt]
\multirow[t]{4}{*}{--} & Source scanning\label{SA1}    & Rule-based & File System traversal \\
                        & Entry detection\label{SA2}    & Rule-based & Annotation match \\
                        & Import closure\label{SA3}     & Rule-based & Breadth-First traversal \\
                        & Bundle assembly\label{SA4}    & Rule-based & File concat \\
--                     & OAS matching\label{SA5}       & LLM        & Cross-file path composition; inheritance chains \\
\texttt{SourceExtractionAgent}                    & Source extraction\label{SA6}  & LLM            & Annotation semantics; inheritance \\
\multirow[t]{2}{*}{\texttt{SemanticExtractAgent}} & Source context compression\label{SA7} & Rule-based     & Fixed schema \\
                                         & Semantic grouping\label{SA8}  & LLM            & Cross-op business logic \\
\hline
\multicolumn{4}{l}{\textit{Oracle Generation}} \\[2pt]
\texttt{SingleOpOracleAgent}                      & Oracle generation\label{OG1}  & LLM            & Path-sensitive reading \\
\texttt{MultiOpOracleAgent}                       & Multi-operation oracles\label{OG2} & LLM            & Multi-step reasoning \\
\texttt{SingleOpChallengerAgent}                  & Review\label{OG3}            & LLM (challenger) & Error detection scales with model \\
\texttt{MultiOpChallengerAgent}                   & Review\label{OG4}            & LLM (challenger) & Error detection scales with model \\
\multirow[t]{2}{*}{\texttt{MastorAgent}}          & Scheduling\label{OG5}         & Rule-based     & Deterministic state machine \\
                                         & Normalization\label{OG6}      & Rule-based     & Fixed filter rules \\
\hline
\end{tabular}
\end{table}

This division operates at two levels.
At the context level, rule-based components produce precisely scoped intermediate representations: the transitive import closure ensures \texttt{SourceExtractionAgent} sees every class affecting an operation's behavior, and the Source context ensures oracle-generation agents reason from verified facts rather than re-interpreting raw source code.
At the quality level, output normalization mechanically removes structurally invalid oracles, and the ChallengerAgent catches semantic errors before they reach the oracle suite (Section~\ref{sec:4.4}).

\subsection{Source Analysis
\label{sec:4.3}}

Before Oracle Generation begins, MASTOR runs the Source Analysis phase that extracts and structures all information oracle agents require.
This phase comprises one rule-based preparation step followed by three LLM-driven extraction steps.

\subsubsection{Rule-based Preparation}
% \man{It may be better to link the description to the subtasks listed in Table~\ref{tab:llm-algo-boundary}.}
% \dsd{Added \textbackslash label\{\} (SA1--SA8, OG1--OG6) to every row in Table~\ref{tab:llm-algo-boundary} and inserted \textbackslash hyperref links at the corresponding sentences throughout Sections~\ref{sec:4.3} and~\ref{sec:4.4}.}
The project source tree is scanned recursively to build a flat file inventory (\hyperref[SA1]{\textit{Source scanning}}).
Entry files are identified by matching file content against a configurable set of framework annotation patterns (\texttt{@RestController} for Spring Boot, while \texttt{@Path} for JAX-RS/Jersey); the matching logic makes no other assumption about the framework, so it extends to additional frameworks by adding patterns without modifying the scanner (\hyperref[SA2]{\textit{Entry detection}}).

For each entry file, a transitive import closure is constructed via breadth-first search traversal of project-internal imports (\hyperref[SA3]{\textit{Import closure}}), and all files in the closure are concatenated with line numbers preserved into a source bundle (\hyperref[SA4]{\textit{Bundle assembly}}).
This ensures that every \texttt{SourceExtractionAgent} instance sees every class whose annotations, field types, or method signatures could influence the operation's behavior.
When multiple bundles share common dependency files, those files are ordered by frequency of appearance to maximize LLM prefix cache utilization across parallel invocations.

% [replaced] \subsubsection{Endpoint Operation Discovery}
% An LLM reads each source bundle and extracts the endpoint operation list:
% HTTP method, complete path template, and entry method name for each operation.
% Although Rule-based Preparation identifies entry files via annotation matching,
% it does not assemble complete path templates: in many frameworks the base path is declared
% at the class level (e.g., \texttt{@RequestMapping("/api/users")}) and the operation-specific
% segment at the method level (e.g., \texttt{@GetMapping("/\{id\}")}), requiring composition across
% two annotation scopes to derive the final template.
% Static string concatenation is unreliable because templates may contain conditional segments,
% inherited base-path declarations, or interface-level annotations;
% the LLM reads the full bundle and resolves this composition from context without requiring
% framework-specific parsing rules.
% The output is the \textbf{endpoint list} that \texttt{MastorAgent} uses to instantiate one \texttt{SourceExtractionAgent} per operation.

\subsubsection{OAS-Anchored Endpoint Matching}
MASTOR uses the project's OAS as the authoritative source of the endpoint operation list.
The OAS file is first parsed algorithmically to extract all declared operations: HTTP method, path template, operationId, parameter declarations, and response schemas.
For each source bundle, the LLM is then given the full OAS operation list and asked to identify which operations are implemented in that bundle, returning the implementing entry class and entry method for each match.
The LLM is explicitly constrained to return only operations from the OAS list; it cannot introduce paths or methods absent from the specification.
Static AST analysis cannot reliably resolve complete path templates, as the final URL is often composed across multiple source locations (e.g., a parent class carries \texttt{@RequestMapping} while the child adds \texttt{@GetMapping}, or a concrete class inherits interface-level annotations without redeclaring them).
The LLM bridges this gap by reasoning about multi-file annotation composition; the OAS list acts as a validation boundary that prevents it from inventing paths.
OAS operations with no matching source method are excluded from the endpoint list and from the mutation score calculation, preventing inflation of reported endpoint coverage.
The output is the endpoint list that \texttt{MastorAgent} uses to instantiate one \texttt{SourceExtractionAgent} per operation; each entry carries the operationId from the OAS for downstream traceability (\hyperref[SA5]{\textit{OAS matching}}).

\subsubsection{Source Context Extraction}
\texttt{MastorAgent} submits one \texttt{SourceExtractionAgent} task per endpoint to the MessageBus;
all tasks run concurrently under the thread-pool limit.
Each agent receives two inputs: the operation's source bundle and the OAS-declared parameter and response schema for that operation.
The LLM reads the source bundle to extract two structured components: \textit{request parameters} (name, binding location, Java type, required flag, source-derived constraints) and \textit{response schema} (success status, body fields with types and nullability, error cases) (\hyperref[SA6]{\textit{Source extraction}}).
It then cross-references each extracted item against the OAS declarations.
Static extraction is insufficient here: Java annotations (\texttt{@PathVariable} vs.\ \texttt{@RequestParam}), inheritance-chain field merging, generic type erasure, and \texttt{ResponseEntity} wrapping all require semantic interpretation that static AST analysis cannot reliably provide.

OAS-declared items that cannot be substantiated in the source code (for example, a status code returned only by a framework-level exception mapper) are placed in separate \texttt{pending} fields rather than in the main Source context.
This precision-biased design prefers to omit an assertion over generating one that cannot be traced to source code, accepting reduced oracle recall in exchange for higher assertion precision.
Downstream oracle-generation agents read only the \texttt{source-verified} components, ensuring that every generated assertion traces to a concrete source code element.
The proportion of OAS-declared items placed in \texttt{pending} and its effect on oracle recall are reported in Section~\ref{sec:6} (RQ3).
Each agent writes its Source context to the OutputStore on completion.
Operations that fail extraction are marked \textsc{Failed} and excluded from oracle generation without halting other operations.
Listing~\ref{lst:source-context-ir} indicates a representative Source context.

% [updated Listing: replaced empty pending arrays with real data from rest-scs output]
\begin{lstlisting}[language=JSON,
caption={Source context for \texttt{GET /api/costfuns/\{i\}/\{s\}} (abridged).
  \texttt{response\_schema\_pending} holds OAS-declared error codes that are not traceable to application-level handlers.},
label={lst:source-context-ir}]
{
  "op_id": "GET /api/costfuns/{i}/{s}",
  "request_params": [
    {"name": "i", "location": "path", "type": "int",
     "required": true, "constraints": [], "oas_match": true},
    {"name": "s", "location": "path", "type": "String",
     "required": true, "constraints": [], "oas_match": true}
  ],
  "response_schema": {
    "success_status": 200,
    "fields": [{"name": "resultAsString", "type": "String",
                "nullable": false}],
    "error_cases": []
  },
  "request_params_pending": [],
  "response_schema_pending": [
    {"status": 401, "oas_claim": "Unauthorized",
     "reason": "returned by framework exception mapper, not application handler"},
    {"status": 403, "oas_claim": "Forbidden",
     "reason": "returned by framework exception mapper, not application handler"},
    {"status": 404, "oas_claim": "Not Found",
     "reason": "returned by framework exception mapper, not application handler"}
  ]
}
\end{lstlisting}

\subsubsection{Semantic Association Extraction}
After all \texttt{SourceExtractionAgent} instances complete,
\texttt{SemanticExtractAgent} processes the full Source context set to identify cross-operation semantic dependencies.
It first compresses each operation's Source context into a compact delimited text format
that fits the full operation set within a single LLM context window (\hyperref[SA7]{\textit{Source context compression}}).
The compression retains all semantically relevant fields while discarding JSON overhead, achieving up to 79\% size reduction (e.g., proxyprint's 68-endpoint set compressed from 109K to 23K characters).
A single LLM call then classifies cross-operation dependencies into five semantic association patterns (Table~\ref{tab:assoc-patterns}), derived from a pilot study on three representative APIs and validated across the remaining 10 benchmark APIs (\hyperref[SA8]{\textit{Semantic grouping}}).
To suppress spurious associations, the prompt enforces two constraints.
First, stateless or purely computational endpoints share no server-side state, so they cannot participate in any sequence.
Second, a sequence is valid only when removing one step causes a later step to fail or lose its business meaning.
The output is a list of operation groups, each containing ordered sequences with
\textit{captures} (named response fields extracted at one step)
and \textit{binds} (parameters wired from earlier captures), expressing cross-step data flow.

\begin{table}[!b]
\centering
\footnotesize
\caption{Semantic Association Patterns Recognized by \texttt{SemanticExtractAgent}}
\label{tab:assoc-patterns}
\setlength{\tabcolsep}{3mm}
\begin{tabular}{lp{8.5cm}}
\hline
\textbf{Pattern} & \textbf{Description} \\
\hline
Resource Lifecycle         & Create, read, update, and delete chains on the same resource type \\
Identifier Flow            & Response identifier consumed as path or body parameter in a later step \\
Field-level Data Passing   & Response field required as input to the next step \\
Nested Resource Dependency & One resource's existence is a prerequisite for another's creation \\
State Propagation          & One operation writes observable server state that another reads \\
\hline
\end{tabular}
\end{table}

\subsection{Oracle Generation
\label{sec:4.4}}

After Source Analysis completes, \texttt{MastorAgent} submits tasks for both the single-op and multi-op paths to the MessageBus.
Both paths execute concurrently; when both complete, \texttt{MastorAgent} merges all oracles from the OutputStore into the oracle suite.

\subsubsection{Single-Operation Oracle Generation (Single-Op Path)
\label{sec:4.4.1}}
\texttt{MastorAgent} submits one oracle generation task per endpoint operation to the MessageBus, and all operations run concurrently.
For each operation, \texttt{SingleOpOracleAgent} reads the Source context from the OutputStore and generates a suite of test oracles around four complementary strategies derived from two orthogonal dimensions: input validity (valid vs.\ invalid) and reasoning direction (forward vs.\ backward) (\hyperref[OG1]{\textit{Oracle generation}}).
Each quadrant in this $2 \times 2$ space is intentionally populated (Table~\ref{tab:oracle-strategies}).

\begin{table}[!t]
\centering
\footnotesize
\caption{Oracle Generation Strategies in \texttt{SingleOpOracleAgent}}
\label{tab:oracle-strategies}
\setlength{\tabcolsep}{2.5mm}
\begin{tabular}{lp{1cm}lp{5.5cm}}
\hline
\textbf{Strategy} & \textbf{Input} & \textbf{Direction} & \textbf{Oracle Target} \\
\hline
Forward-Valid (FV)    & Valid   & Input $\to$ response &
  Assert that all non-nullable fields declared in the response schema are present and non-null;
  assert that typed fields have the expected type \\
Forward-Invalid (FI)  & Invalid & Input $\to$ response &
  Assert that the error status code matches the error case declared in the Source context
  for the triggered condition \\
Backward-Valid (BV)   & Valid   & Source context $\to$ input &
  Derive multiple valid input combinations from the declared parameter constraints,
  covering distinct boundary conditions, to increase valid-path diversity \\
Backward-Invalid (BI) & Invalid & Error case $\to$ input &
  Derive the specific invalid input that matches each declared error condition,
  by reasoning backward from the condition description to a concrete input value \\
\hline
\end{tabular}
\end{table}

The FV and FI strategies share a set of path-sensitive reasoning rules encoded in the LLM prompt:
for each numeric or length constraint $c$, the agent instantiates inputs at $c-1$, $c$, and $c+1$
to straddle the validity boundary (boundary-value enumeration);
for each branching condition, the agent produces at least one oracle per branch direction
(branching-pair coverage);
for exception-handling blocks, the agent reads catch-clause types and maps them to
the corresponding error status codes (catch-block analysis).
FV and FI reason forward from input to response and require branch tracing; BV and BI reason backward from the Source context to input, so there is no branching structure to enumerate.
These rules are encoded in the LLM prompt, exploiting the LLM's ability to focus on semantically relevant branches without exhaustive control flow graph (CFG) enumeration.
After generation, a deterministic normalization step applies three filter rules (\hyperref[OG6]{\textit{Normalization}}).
First, any oracle that carries no status assertion is discarded, as a status assertion is the minimal verifiable claim.
Second, any FV or BV oracle that carries no field assertion is discarded, because a success-case oracle that only checks the status code cannot detect faults in the response body.
Third, any input key outside the four recognized binding locations (\texttt{path}, \texttt{query}, \texttt{body}, and \texttt{headers}) is stripped from the input object, eliminating hallucinated parameter names introduced by the LLM without discarding the oracle itself.

\begin{table}[!b]
\centering
\footnotesize
\caption{Review Dimensions Evaluated by Challenger Agents}
\label{tab:review-dimensions}
\setlength{\tabcolsep}{2.6mm}
\begin{tabular}{lp{10cm}}
\hline
\textbf{Dimension} & \textbf{Criteria} \\
\hline
Structural completeness &
  Body \texttt{not\_null} assertion present on all successful cases;
  every non-nullable field in the response schema is covered \\
Coverage gaps &
  Boundary values at $c \pm 1$; both branching directions exercised;
  at least three distinct valid input combinations \\
Assertion correctness &
  Field names and status codes match the source context;
  no hallucinated identifiers or impossible conditions \\
Sequence logic &
  (Multi-op only) Captures reference declared names;
  binds match declared input parameters;
  state transitions are semantically coherent \\
\hline
\end{tabular}
\end{table}

The normalized oracle suite is then passed to \texttt{SingleOpChallengerAgent} together with the Source context (\hyperref[OG3]{\textit{Review}}).
\texttt{SingleOpChallengerAgent} evaluates the suite along three dimensions (Table~\ref{tab:review-dimensions}).
\texttt{SingleOpChallengerAgent} outputs natural-language hints without producing corrected oracles.
If hints are non-empty, \texttt{SingleOpOracleAgent} performs one targeted regeneration pass guided by those hints, revising only the flagged oracles and preserving all others.
In the default configuration, oracle generation uses DeepSeek V4 Pro as the primary model and Qwen3.6-Plus~\cite{qwen} as the challenger.
The verified oracles are written to the OutputStore under the operation's identifier.
Listing~\ref{lst:oracle-ir} indicates representative oracles produced by the single-op path.

\begin{lstlisting}[language=JSON,
caption={Oracle entries for \texttt{GET /api/costfuns/\{i\}/\{s\}} (Single-Op Path, rest-scs).
  The FI entry asserts a type-mismatch 400; the FV entry traces the result value through sequential branch logic.},
label={lst:oracle-ir}]
[
  {
    "test_id": "fi_i_not_integer",
    "description": "FI: i is not an integer triggers 400",
    "input": {"path": {"i": "abc", "s": "test"}, "query": {}},
    "assertions": [{"type": "status", "expected": 400}],
    "evidence": "Spring cannot convert 'abc' to Integer for path variable i, resulting in 400 Bad Request.",
    "oracle_strategy": "fi"
  },
  {
    "test_id": "fv_i_5_s_baab",
    "description": "FV: i=5, s='baab' exercises sequential branch logic",
    "input": {"path": {"i": "5", "s": "baab"}, "query": {}},
    "assertions": [
      {"type": "status", "expected": 200},
      {"type": "field", "field_path": "", "op": "not_null", "expected": null},
      {"type": "field", "field_path": "", "op": "equals", "expected": "10"}
    ],
    "evidence": "Costfuns.subject: i==5 sets result=1; s.equals('ba'+'ab') sets result=7; s != s2+s2 sets result=10. Sequential ifs overwrite.",
    "oracle_strategy": "fv"
  }
]
\end{lstlisting}

The \texttt{evidence} field in each oracle is a natural-language annotation produced by the oracle-generation agent, citing the specific class, method, or code path in the source bundle that justifies the assertion.
It is not a structured reference (e.g., a file path and line number), but it serves as a human-readable audit trail linking each assertion to the source code element from which it was derived.
This grounding is reinforced at two levels: the normalization step discards oracles that carry no verifiable claim and strips structurally invalid inputs, and the ChallengerAgent's assertion-correctness dimension explicitly checks that all field names and status codes in the oracle suite match the verified source context rather than the LLM's prior knowledge.

\subsubsection{Multi-Operation Oracle Generation (Multi-Op Path)}
The single-op path treats each endpoint operation in isolation and cannot detect failures in cross-operation contracts.
The multi-op path addresses this gap by synthesizing multi-operation oracles grounded in the semantic associations
identified by \texttt{SemanticExtractAgent}.

The multi-op path consumes the operation groups produced by \texttt{SemanticExtractAgent}.
If no groups were found, the multi-op path produces no oracles and terminates immediately.
Otherwise, \texttt{MastorAgent} submits one oracle generation task per group to the MessageBus,
and all groups run concurrently.
We report the frequency and group-size distribution of triggered multi-op groups across the 13 subject APIs in Section~\ref{sec:6} (RQ1).

For each group, \texttt{MultiOpOracleAgent} receives the ordered operation sequences together with their source contexts and generates multi-operation oracles (\hyperref[OG2]{\textit{Multi-operation oracles}}) in which the LLM instantiates concrete test inputs for each step, either directly from parameter constraints or by binding values captured from prior steps' responses. 
% \man{it is a bit unclear whether the approach generates test inputs.}
% \dsd{Rewritten to explicitly state that the LLM instantiates test inputs, either directly from parameter constraints or via captures/binds from prior steps.}
Data flow between steps is made explicit via two constructs:
\textit{captures}, which name response fields to be extracted at one step,
and \textit{binds}, which wire those captured values into the input parameters of a later step.
For example, a \texttt{POST /users} step may capture the returned \texttt{id} field,
which is then bound to the \texttt{\{userId\}} path parameter of a subsequent \texttt{GET /users/\{userId\}} step.
This explicit encoding prevents the LLM from treating each step as an independent request and forces reasoning about inter-step state transitions.

Status and field assertions are generated for each step.
Assertions on later steps may reference captured values via \texttt{\$captures.<name>} syntax,
enabling the oracle to verify not only that each step succeeds individually
but that the response of one step is consistent with the input accepted by the next.

After generation, the same deterministic normalization step and review-and-regeneration step described for the single-op path (Section~\ref{sec:4.4.1}) are applied,
with \texttt{MultiOpChallengerAgent} (\hyperref[OG4]{\textit{Review}}) giving particular attention to the Sequence logic dimension,
which checks that captures and binds are grounded in declared field names and input parameters
and that the overall state transition is semantically coherent.
Verified multi-operation oracles are written to the OutputStore under their sequence identifiers.

% \subsection{Implementation}
% MASTOR is implemented in Python 3.12.
% The orchestrator and shared infrastructure (OutputStore, MessageBus, budget tracker) together comprise approximately 1,200 lines of code.
% All seven agents share a common base class that provides logging, timing, and LLM invocation utilities;
% each agent's task-specific prompt construction and response parsing logic adds 150--400 lines,
% keeping individual agents compact and independently modifiable.
% LLM calls are made through a thin abstraction layer that supports multiple providers via OpenAI-compatible APIs;
% token usage and cost are tracked per phase to support the RQ4 efficiency analysis.
% The oracle conversion tools--\textit{toJUnit} and \textit{toPostmanAssertify}--are implemented as standalone Python modules totaling approximately 800 lines.
% \textit{toJUnit} translates the oracle suite into JUnit 4 test cases with RestAssured assertions,
% suitable for execution against a locally deployed SUT during mutation testing.
% \textit{toPostmanAssertify} translates the oracle suite into Postman test scripts for CI/CD integration.
% All experiments are orchestrated through a command-line entry point that manages project configuration,
% phase scheduling, and artifact export.

\section{Experimental Methodology
\label{sec:5}}
This section describes the experimental methodology, including the research questions, experimental setup, evaluation metrics, and experimental procedures.

\subsection{Research Questions
\label{sec:5.1}}
To comprehensively evaluate the effectiveness, robustness, and practicality of MASTOR, we investigate the following research questions:

\textbf{RQ1:} How effective are semantic test oracles generated by MASTOR at detecting faults?

This RQ evaluates MASTOR's overall oracle performance by measuring the mutation score achieved by the oracle suite across all 13 subject APIs.

\textbf{RQ2:} Compared with Direct Prompting and SATORI, how do MASTOR-generated oracles perform?

This RQ evaluates MASTOR against Direct Prompting and SATORI~\cite{alonsoSATORIStaticTest2025a} on 50 selected endpoint operations under identical experimental conditions.

\textbf{RQ3:} How do MASTOR’s key architectural components contribute to oracle quality and performance?

This RQ quantifies the individual contribution of two architectural components (\textit{Multi-Op Oracle Generation} and \textit{ChallengerAgent}) by evaluating three ablated variants against Full MASTOR on all 13 subject APIs.

\textbf{RQ4:} What is the computational cost of MASTOR oracle generation?

This RQ measures token consumption, inference time, and monetary cost across all 13 subject APIs and analyzes cost variation in relation to project characteristics such as endpoint count and codebase size.

\subsection{Experimental Setup
\label{sec:5.2}}

\subsubsection{Subject APIs and LLMs}
To select representative subject APIs, we surveyed the SUT projects used in over 60 RESTful API testing studies and ranked them by frequency of appearance.
We retained all projects appearing in the top five frequency tiers (i.e., projects referenced by at least eight studies), yielding 13 subject APIs.
Table~\ref{TAB:APIs} lists all 13 subject APIs with their characteristics.
The 13 subject APIs span a wide range in scale: from \texttt{rest-ncs} (605 LoC, 6 endpoints) to \texttt{languagetool} (174,781 LoC, 2 endpoints) by code size, and to \texttt{proxyprint} (8,338 LoC, 68 endpoints) by endpoint count.
All projects are implemented in Java; nine use Spring Boot, two use Jersey, and two use JDK-based HTTP servers.
The source code of all subject APIs, the generated oracle suite, and the full experimental data are available at~\cite{mastorGitHub}.
The benchmark totals 296 endpoint operations across 251,303 lines of code.
Table~\ref{TAB:coverage} summarizes the coverage of the 13 subject APIs across five key dimensions.

\begin{table}[!t]
\centering
\footnotesize
\caption{Summary of 13 Subject RESTful APIs Ranked by Appearance Frequencies in Previous Studies}
\label{TAB:APIs}
\setlength{\tabcolsep}{5.1mm}
\begin{tabular}{lrrrrl}
\hline
\textbf{API} & \textbf{Tier} & \textbf{\#Count} & \textbf{LoC} & \textbf{\#Endpoint} & \textbf{Framework} \\
\hline
restcountries~\cite{fayderRestcountries}$^{\ddagger}$           & 1 & 15 & 1,977   & 22 & Jersey      \\
features-service~\cite{martinezFeaturesService}$^{\ddagger}$   & 2 & 13 & 2,275   & 18 & Jersey      \\
languagetool~\cite{languagetoolGitHub}                          & 2 & 13 & 174,781 & 2  & JDK         \\
genome-nexus~\cite{genomeNexusGitHub}                           & 3 & 12 & 30,004  & 23 & Spring Boot \\
market~\cite{marketGitHub}$^{\ddagger}$                         & 4 & 9  & 9,861   & 13 & Spring Boot \\
rest-ncs~\cite{arcuriAdvancedWhiteBoxHeuristics2024}            & 4 & 9  & 605     & 6  & Spring Boot \\
rest-scs~\cite{arcuriAdvancedWhiteBoxHeuristics2024}            & 4 & 9  & 862     & 11 & Spring Boot \\
proxyprint~\cite{proxyprintGitHub}$^{\dagger}$                  & 5 & 8  & 8,338   & 68 & Spring Boot \\
catwatch~\cite{catwatchGitHub}$^{\dagger}$                      & 5 & 8  & 9,636   & 14 & Spring Boot \\
person-controller~\cite{personControllerGitHub}                 & 5 & 8  & 1,112   & 12 & Spring Boot \\
user-management~\cite{userManagementGitHub}                     & 5 & 8  & 4,274   & 21 & Spring Boot \\
tracking-system~\cite{trackingSystemGitHub}                     & 5 & 8  & 5,947   & 67 & Spring Boot \\
swagger-petstore~\cite{swaggerPetstoreGitHub}                   & 5 & 8  & 1,631   & 19 & JDK         \\
\hline
\textbf{\textit{Total}} & & & \textbf{251,303} & \textbf{296} & \\
\hline
\end{tabular}
\par
\begin{minipage}{\linewidth}
\textit{Note.}
\#Count = number of prior REST API testing studies referencing the API.
Tier = frequency tier derived from \#Count (Tier~1 = most frequently referenced).
\#Endpoint = number of endpoint operations after excluding framework-internal endpoints.
$^{\dagger}$ proxyprint and catwatch expose framework management endpoints auto-generated by Spring Boot Actuator (/health, /metrics, /env, /mappings, etc.) that are not application-logic endpoints.
The \#Endpoint values shown for proxyprint (68) and catwatch (14) are post-exclusion counts (47 and 9 endpoints excluded, respectively).
$^{\ddagger}$ Pilot APIs used to derive the five semantic association types (Section~\ref{sec:4.3}); the remaining 10 APIs serve as the held-out validation set.
\end{minipage}
\end{table}

\begin{table}[!t]
\centering
\footnotesize
\caption{Coverage of 13 Subject APIs Across Five Benchmark Dimensions}
\label{TAB:coverage}
\setlength{\tabcolsep}{3mm}
\begin{tabular}{lp{11cm}}
\hline
\textbf{Dimension} & \textbf{Coverage} \\
\hline
Endpoint scale & 2 $\rightarrow$ 6 $\rightarrow$ 11 $\rightarrow$ 12 $\rightarrow$ 13 $\rightarrow$ 14 $\rightarrow$ 18 $\rightarrow$ 19 $\rightarrow$ 21 $\rightarrow$ 22 $\rightarrow$ 23 $\rightarrow$ 67 $\rightarrow$ 68 \\
Database       & None (5 APIs) $\rightarrow$ H2 (4) $\rightarrow$ MySQL (2) $\rightarrow$ PostgreSQL (2) $\rightarrow$ MongoDB (2) \\
Framework      & Spring Boot (9 APIs) + Jersey (2) + JDK (2) \\
Domain         & Numerical computation, string computation, text grammar checking, e-commerce, analytics dashboard, feature management, RBAC authorization, data query, bioinformatics, project tracking, print service, personnel management, pet store management \\
Seed data      & 7 self-contained (in-memory, embedded JSON, or embedded H2) + 6 requiring pre-database setup \\
\hline
\end{tabular}
\end{table}

For oracle generation, we use DeepSeek V4 Pro~\cite{deepseekDeepSeek} as the primary LLM and Qwen3.6-Plus~\cite{qwen} as the challenger LLM. 
%\wang{What is the rationale for selecting these two models? I think it should be clearly stated. What would be the result if Qwen3.6-Plus is used as the primary LLM and DeepSeek as the challenger LLM?}
%\dsd{Added rationale in the following paragraphs. The sensitivity to model assignment is discussed in Section~\ref{sec:7}.}
The two roles impose different capability requirements: the primary LLM performs oracle synthesis from implementation source code, requiring strong code-reasoning capability; the challenger LLM performs a single-pass review to identify logical weaknesses, where instruction-following precision is the dominant concern.
DeepSeek V4 Pro achieves 93.5\% on LiveCodeBench and a Codeforces rating of 3,206~\cite{deepseekV4ProHF}, matching or exceeding proprietary models such as GPT-5.4 and Gemini-3.1 Pro on code-reasoning benchmarks, making it the stronger candidate for oracle synthesis.
Qwen3.6-Plus offers competitive general reasoning (GPQA Diamond: 86.0\%)~\cite{qwen36HF} at lower per-token cost, making it well-suited for the challenger role where cost scales with oracle count.

Both models are open-source and accessed via API at substantially lower cost than proprietary alternatives such as GPT or Claude, making the approach reproducible without high inference budgets.
The open-source nature of both models also supports local deployment, which is relevant in security-sensitive contexts where transmitting proprietary source code to an external API is not acceptable.
All prompts and hyperparameters are held constant across all experiments.

\subsubsection{Evaluation Metrics
\label{sec:5.3}}
% \man{There may be a potential risk since F1 score and precision are not included in the evaluation. However, we can also wait for the reviewers' feedback.}
% \dsd{Added justification for not using F1/precision: a complete ground-truth oracle set does not exist for implementation-derived oracles, making precision and recall undefined; MS is the established metric for oracle quality evaluation.}
For each research question, we use the following evaluation metrics, listed as follows:
\begin{itemize}
    \item \textbf{Mutation Score (MS).} MS serves as the primary evaluation metric for RQ1, RQ2, and RQ3, quantifying the fault-detection capability of the generated oracles through mutation testing.
A source code mutant $m$ is generated by applying a single mutation operator to the subject API's Java source code.
For each mutant, the mutated API is compiled, deployed, and exercised by the test input suite.
A mutant is \textit{killed} if at least one oracle assertion fails on any test response produced by the mutated API. 
The mutation score is defined as:
\begin{equation}
    \mathit{MS} = \frac{|M_{\mathit{killed}}|}{|M_{\mathit{killed}}| + |M_{\mathit{survived}}|}, 
\end{equation}
where $M_{\mathit{killed}}$ is the set of mutants killed by at least one oracle assertion
and $M_{\mathit{survived}}$ is the set of mutants that no assertion detected.
PITest's \textsc{no\_coverage} mutants (those not reached by any test input) are excluded from both sets.
Equivalent mutants are not explicitly identified; surviving mutants may include some equivalent ones,
which is discussed as a threat to validity in Section~\ref{sec:threats}.
For RQ1 and RQ3, MS is computed from the full oracle suite produced by MASTOR running across all endpoint operations of each subject API, comprising status, field, and multi-operation oracle types.
For RQ2, the three tools under comparison (MASTOR, Direct Prompting, and SATORI) are all evaluated on an identical set of 50 selected endpoint operations,
and MS is restricted to status and field oracle types to match the oracle types that SATORI can produce; the absolute killed count per operation is also reported alongside MS.
MS is adopted as the primary metric in preference to precision- and recall-based alternatives such as F1-Score, because no complete ground truth oracle set exists for implementation-derived semantic oracles: the full set of correct assertions for an arbitrary API implementation cannot be enumerated without exhaustive manual specification.
MS measures fault detection capability directly without requiring a ground truth, and is the standard evaluation metric for oracle quality in the literature~\cite{barrOracleProblemSoftware2015,jiaAnalysisMutationTesting2011}.

\item \textbf{Oracle Count.}
For RQ1 and RQ3, we report the number of oracles generated by types (\#Status, \#Field, and \#Multi-Op) and the total number of oracles (\#Oracles).
For RQ2, we report the killed mutant count per selected operation alongside MS, enabling per-operation comparison across the three tools.
Oracle counts characterize the breadth of the generated oracle suite and complement MS by showing how coverage distributes across oracle types.

\item \textbf{Cost Metrics.}
For RQ4, we record four cost dimensions per API: the number of LLM calls (\#Calls), token consumption (input and output tokens separately, in millions), total wall-clock inference time (seconds), and estimated monetary cost (USD) based on each provider's published per-token pricing.
All four dimensions are reported both in aggregate and broken down by agent (Endpoint Discovery, Source Extraction, Single-Op Generation, Single-Op Challenger, Multi-Op Generation, Multi-Op Challenger).
Per-operation cost is computed as total cost divided by the number of endpoint operations, enabling normalized comparison across APIs of different scales.
\end{itemize}

\subsection{Experimental Procedure
\label{sec:5.4}}

\begin{figure}[!t]
  \centering
  \includegraphics[width=\textwidth]{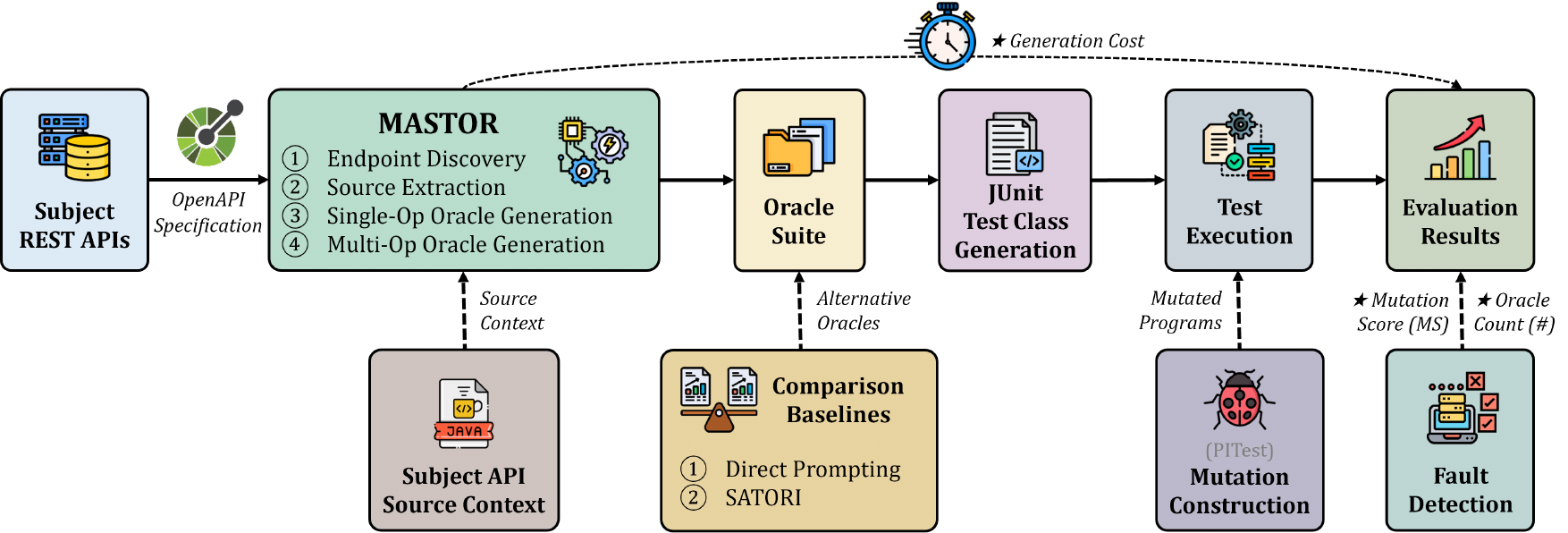}
  \caption{Overview of the experimental procedure. $\bigstar$-marked labels denote evaluation metrics: Mutation Score (MS) and Oracle Count (\#, number of oracles) on the fault-detection-to-results path for RQ1--RQ3; the dashed arc from MASTOR to the evaluation results represents generation cost for RQ4.}
  \Description{A flowchart of the experimental procedure. The main pipeline runs left to right: Subject REST APIs feed into MASTOR, producing an oracle suite, which is converted into JUnit test classes and executed via PITest, yielding Evaluation Results for RQ1 through RQ4. Supporting components include Subject API source context, Comparison Baselines consisting of Direct Prompting and SATORI, Mutation Construction via PITest, and Fault Detection. A dashed arc from MASTOR to the evaluation results represents the generation cost measured for RQ4.}
  \label{fig:exp-procedure}
\end{figure}

Figure~\ref{fig:exp-procedure} presents the basic experimental procedure.
The main pipeline proceeds as follows.
13 subject REST APIs are fed into MASTOR, which performs Endpoint Discovery, Source Extraction, Single-Op Oracle Generation, and Multi-Op Oracle Generation using the per-API source context to produce an oracle suite.
The oracle suite is converted into JUnit test classes and then executed against PITest, which applies mutation operators to the subject API bytecode and records whether each oracle assertion detects the induced fault, yielding a Mutation Score reported in the Evaluation Results.
For RQ2, two comparison baselines---Direct Prompting (DP) and SATORI---provide alternative oracles that replace the MASTOR-generated output within the same execution pipeline, enabling side-by-side mutation score comparison.
The dashed arc from MASTOR to the evaluation results represents generation cost (token consumption and wall-clock inference time) measured across the MASTOR generation stage and reported for RQ4.
Mutation Score (MS) and Oracle Count (\#, number of oracles), labeled on the path from fault detection to the Evaluation Results, are the primary evaluation metrics for RQ1, RQ2, and RQ3.
All three metrics are marked with $\bigstar$ in the figure.

\subsubsection{Shared Setup}
The shared setup includes mutant generation, input value alignment, and oracle conversion into JUnit test classes, all applied uniformly across all four RQs.

\textbf{Mutant Generation.}
We use the PITest Maven Plugin~\cite{colesPITPracticalMutation2016} to generate mutants for each subject API.
The \textsc{stronger} mutator group is applied, which comprises twelve mutation operators:
Conditionals Boundary, Increments, Invert Negatives, Math, Negate Conditionals,
Void Method Calls, Empty Returns, False Returns, True Returns, Null Returns,
Primitive Returns, and Experimental Switch.
Each mutant introduces exactly one mutation into the Java bytecode.
PITest's mutant generation statistics for each subject API, including killed, survived, and no-coverage counts, are reported in Table~\ref{TAB:RQ1}. 
%{\wang{How many mutants were generated by this tool for each experimental project, and how many mutants were discarded? I suggest reporting the above experimental data and analyzing the main reasons why mutants were discarded or could not be loaded.}} \dsd{I made a mistake in this sentence and have corrected it. Pitest performs mutations at the bytecode level and will not discard. Sorry.}
PITest categorizes each surviving mutant as \textsc{killed}, \textsc{survived}, or \textsc{no\_coverage};
only \textsc{killed} and \textsc{survived} mutants are included in the mutation score denominator.

\textbf{Oracle Suite Execution.}
Running the generated oracle suites in PITest requires two steps: aligning test input values with the actual database state, and converting oracle suites into JUnit test cases.

For input value alignment, many endpoint operations require database-resident resources to produce non-empty, assertable responses.
A \texttt{GET /users/\{id\}} oracle asserting a specific field value is only exercisable if a user with that \textit{id} exists in the database at test time; otherwise, the API returns HTTP status code \texttt{404} and all field assertions fail silently, contributing nothing to mutation detection.
The 13 subject APIs fall into three categories based on their data initialization mechanism, as summarized in Table~\ref{TAB:input-align}.
\textit{Self-contained} APIs carry all data within the deployment artifact, or are embedded in an in-memory database, so LLM-generated input values are inherently valid.
\textit{Seeded external-DB} APIs populate their database automatically via deterministic, version-controlled seed scripts; MASTOR reads those scripts during source context extraction and constrains the LLM to generate input values drawn from the known seed records.
\textit{Externally-dependent} APIs require an external database service that does not auto-populate on startup; for these APIs, we start the required database in a Docker container and inject initialization data programmatically via JUnit \texttt{@BeforeClass} lifecycle hooks in the generated test class, ensuring that the required records exist before any oracle assertion executes.

% For the oracle conversion, we implement three Python-based tools that consume the MASTOR-generated oracle suite (see Listing~\ref{lst:oracle-ir}; full oracle suite for all 13 subject APIs available at~\cite{mastorGitHub}).
For the oracle conversion, we implement three Python-based tools.
Listing~\ref{lst:oracle-ir} indicates the MASTOR-generated oracle suite they consume, the full oracle suite for all 13 subject APIs available at~\cite{mastorGitHub}.
% that consume the MASTOR-generated oracle suite (see Listing~\ref{lst:oracle-ir}; full oracle suite for all 13 subject APIs available at~\cite{mastorGitHub}).
All three tools use \texttt{op\_id} and \texttt{o\_id} as unique identifiers, enabling cross-tool traceability for individual oracles.
a) \texttt{ToJUnit} converts the oracle suite into JUnit test cases executed by the PITest Maven Plugin; mutation score for RQ1, RQ2, and RQ3 is computed from these JUnit executions, and a mutant is killed if at least one assertion fails on the mutated API's response.
b) \texttt{ToPostmanAssertify} converts the oracle suite into Postman test scripts compatible with PostmanAssertify~\cite{alonsoAGORAAutomatedGeneration2023}, an open-source framework for executing Postman-format API test assertions, for practical deployment use.
c) \texttt{ToReadable} converts the oracle suite into human-readable oracle descriptions, serving as an inspection aid for manual verification of oracle correctness.

\begin{table}[!t]
\centering
\footnotesize
\caption{Input Value Alignment Classification for 13 Subject APIs}
\label{TAB:input-align}
\setlength{\tabcolsep}{7mm}
\begin{tabular}{lll}
\hline
\textbf{API} & \textbf{Database} & \textbf{Input Value Source} \\
\hline
\multicolumn{3}{l}{\textit{Self-contained (7 APIs)}} \\
\quad rest-ncs         & None          & Pure computation \\
\quad rest-scs         & None          & Pure computation \\
\quad restcountries    & None          & Classpath JSON \\
\quad languagetool     & None          & Pure computation \\
\quad swagger-petstore & None          & Auto-generated examples \\
\quad features-service & H2 (embedded) & \texttt{data.sql} (auto on startup) \\
\quad market           & H2 (embedded) & \texttt{data.sql} (auto on startup) \\
\hline
\multicolumn{3}{l}{\textit{Seeded external-DB (2 APIs)}} \\
\quad user-management  & MySQL         & \texttt{data.sql} / \texttt{schema.sql} (read by MASTOR) \\
\quad tracking-system  & H2 / MySQL    & Flyway V1--V18 migrations (read by MASTOR) \\
\hline
\multicolumn{3}{l}{\textit{Externally-dependent (4 APIs)}} \\
\quad proxyprint        & PostgreSQL      & \texttt{@BeforeClass} injection (Docker container) \\
\quad catwatch          & H2 / PostgreSQL & \texttt{@BeforeClass} injection (Docker container) \\
\quad genome-nexus      & MongoDB         & \texttt{@BeforeClass} injection (Docker container) \\
\quad person-controller & MongoDB         & \texttt{@BeforeClass} injection (Docker container) \\
\hline
\end{tabular}
\par
\begin{minipage}{\linewidth}
\textit{Note.}
H2 (embedded) = H2 in-memory database started within the same JVM process.
For seeded external-DB APIs, MASTOR reads the seed scripts at source context extraction time to constrain generated input values to known records.
For externally-dependent APIs, test data is injected programmatically in the JUnit \texttt{@BeforeClass} setup, with the external database started in a Docker container.
\end{minipage}
\end{table}

\subsubsection{Per-RQ Procedures}

\textbf{Oracle Generation (RQ1).}
We run MASTOR on all 13 subject APIs to generate the full oracle suite, comprising status, field, and multi-operation oracle types without any type restriction.
All generated oracles participate in PITest; we collect oracle counts by type and MS per API.

\textbf{Baseline Comparison (RQ2).}
We compare three tools on 50 selected endpoint operations drawn from the 13 subject APIs: MASTOR, Direct Prompting (DP), and SATORI~\cite{alonsoSATORIStaticTest2025a}.
AGORA+~\cite{alonsoAGORAAutomatedGeneration2023} is excluded because it derives oracles from execution traces rather than LLM-based inference; all three compared tools share the same LLM-driven oracle generation paradigm, ensuring a methodologically coherent comparison.
For each subject API, we select up to four endpoint operations, aiming to cover diverse HTTP methods.
\texttt{languagetool} is the only exception, contributing two operations as its OAS declares only two endpoint operations in total (not a selection choice on our part), yielding 50 selected operations in total.
% \man{this setting may be questionable. however, we could also wait for reviewers' comments.}
% \dsd{Clarified that languagetool has only 2 endpoints in its OAS, so the reduced count is an API-level constraint, not a sampling bias. Note: we can run a full-set RQ2 experiment (all endpoints per API) if you consider it necessary; the current setup was constrained by time.}

DP is a zero-shot baseline that issues a single LLM call with the OAS-defined operation list and the concatenated source bundle, without any multi-agent coordination, structured intermediate representations, or chain-of-thought guidance.
% The full prompt template is provided in Listing~\ref{lst:dp-prompt} in Appendix~\ref{appendix:dp-prompt}.
% Listing \ref{lst:dp-prompt} indicates the structure of the full prompt template.
Listing \ref{lst:dp-prompt} indicates the zero-shot prompt template used by the Direct Prompting (DP).

All three tools are restricted to status and field oracle types; MASTOR and DP filter their full outputs to the 50 selected operations, while SATORI generates oracles for those operations directly.
Source code mutants from RQ1 are reused; a mutant is killed when at least one oracle assertion detects a behavioral deviation in the API response.
We report MS and killed count per operation for all three tools.

\textbf{Ablation Study (RQ3).}
We evaluate three ablated configurations against Full MASTOR on all 13 subject APIs:
\begin{itemize}
  \item \textit{w/o Multi-Op Oracle Generation}: The multi-op path is disabled; only single-op oracles are generated.
  \item \textit{w/o ChallengerAgent}: The review-and-regeneration step is skipped; first-pass oracle output is used directly.
  \item \textit{w/o Both}: Both components are disabled simultaneously, serving as the minimal baseline.
\end{itemize}
We report MS and oracle count for each configuration; $\Delta$MS denotes the absolute MS change relative to Full MASTOR.

\textbf{Efficiency Analysis (RQ4).}
We record the computational cost for each of the 13 subject APIs.
Token consumption (input and output separately) and wall-clock inference time are logged at each phase boundary:
Source Analysis (Ph-A), single-operation oracle generation (Ph-B-SO), multi-operation oracle generation (Ph-B-MO), and ChallengerAgent review (Ph-B-R).
Monetary cost is computed post hoc from published per-token pricing.
We report per-API and per-operation cost metrics to characterize cost variation across projects.

\section{Experimental Results
\label{sec:6}\label{sec:evaluation}}
In this section, we present the experimental results to answer the above four RQs.
Throughout this section, MS denotes the mutation score as defined in Section~\ref{sec:5.3}.
For RQ1 and RQ3, MS is computed from MASTOR's full oracle output across all endpoint operations.
For RQ2, MS is computed from oracles restricted to the 50 selected operations and to status and field types, and the killed count per operation is also reported.

\subsection{Answer to RQ1: How Effective Are Semantic Test Oracles Generated by MASTOR at Detecting Faults?}
\label{sec:6.1}

Table~\ref{TAB:RQ1} reports, for each subject API, the oracle counts by type and the mutation results achieved by the MASTOR oracle suite.

\begin{table*}[!t]
\centering
\footnotesize
\caption{Oracle Counts by Types and Mutation Testing Results for Each Subject API}
\label{TAB:RQ1}
\setlength{\tabcolsep}{2.7mm}
\begin{tabular}{l | rrrr | rrrr}
\hline
\multirow{2}{*}{\textbf{API}}
  & \multicolumn{4}{c|}{\textbf{Oracles}}
  & \multicolumn{4}{c}{\textbf{Mutants}} \\
\cline{2-9}
  & \textit{\#Status} & \textit{\#Field} & \textit{\#Multi-Op} & \textit{\#Oracles}
  & \textit{Killed} & \textit{Survived} & \textit{No\_Cov} & \textit{MS (\%)} \\
\hline
restcountries     &    221 &    108 & 18 &     365 &  301 &    13 &    48 & 95.9 \\
features-service  &    402 &    272 & 19 &     693 &  296 &    65 &    25 & 82.0 \\
languagetool      &     18 &      8 &  0 &      26 &    8 &     3 & 1,318 & 72.7 \\
genome-nexus      &  1,062 &    628 & 31 &   1,721 &  648 &   252 &   800 & 72.0 \\
market            &    457 &    269 & 30 &     756 &  468 &   182 &   250 & 72.0 \\
rest-ncs          &     95 &     66 &  0 &     161 &  435 &    89 &    15 & 83.0 \\
rest-scs          &    204 &    196 &  0 &     400 &  241 &    74 &    31 & 76.5 \\
proxyprint        &    618 &    362 & 40 &   1,020 &  284 &   111 &   788 & 71.9 \\
catwatch          &    430 &    228 & 11 &     669 &  342 &   108 &   150 & 76.0 \\
person-controller &    848 &    549 & 31 &   1,428 &  525 &   175 &   100 & 75.0 \\
user-management   &    622 &    322 & 42 &     986 &  638 &   212 &   150 & 75.1 \\
tracking-system   &    964 &    466 & 27 &   1,501 &  690 &   310 &   500 & 69.0 \\
swagger-petstore  &    161 &    101 & 14 &     296 &  188 &    61 &    81 & 75.5 \\
\hline
\textbf{\textit{Total}}
  & 6,102 & 3,575 & 263 & 10,022 & 5,064 & 1,655 & 4,256 & 75.4\% \\
\hline
\end{tabular}
\par
\begin{minipage}{\linewidth}
\textit{Note.}
\#Status, \#Field, \#Multi-Op = number of oracle entries whose primary assertion type is status, field, or multi-operation, respectively.
A single oracle entry may carry both a status and field assertion; such entries are counted in \#Status and may also be reflected in \#Field, so the column sum may exceed \#Oracles.
\#Oracles = total number of oracle entries for the API.
MS = mutation score (\%) = Killed / (Killed + Survived); No\_Cov mutants are excluded from both sets.
\end{minipage}
\end{table*}

MASTOR generates 10,022 oracles across all 13 subject APIs, averaging 771.0 oracles per API.
Status oracles account for the largest share (6,102, 60.9\%), reflecting MASTOR's ability to synthesize precise HTTP status assertions from source code evidence.
Field oracles contribute 3,575 (35.7\%), demonstrating MASTOR's capacity to trace response field provenance through the call graph.
Multi-operation oracles contribute 263 (2.6\%) and are concentrated in APIs with complex cross-resource dependencies (e.g., \texttt{proxyprint}, \texttt{tracking-system}); stateless APIs such as \texttt{rest-ncs} and \texttt{rest-scs} produce none.

Across all 13 APIs, the oracle suite achieves an overall MS of 75.4\% (5,064 killed out of 6,719 covered mutants), with individual API scores ranging from 69.0\% to 95.9\%.
The results follow a clear pattern driven by two factors: response predictability and code reachability.
APIs with static or purely computed responses score highest: \texttt{restcountries} reaches 95.9\% by asserting well-defined fields drawn directly from a classpath dataset, and \texttt{rest-ncs} reaches 83.0\% because its numerical computation endpoints (e.g., \texttt{remainder}, \texttt{gammq}) yield values directly derivable from source code.
APIs with simple CRUD semantics and deterministic seed data cluster in the 75\%--82\% range (e.g., \texttt{features-service} at 82.0\%, \texttt{user-management} at 75.1\%).
APIs with authentication barriers or deeply nested response structures score in the 69\%--73\% range.
\texttt{languagetool} records a No\_Cov of 1,318 because the grammar-checking engine contains extensive rule-processing code unreachable by HTTP-level inputs.
\texttt{proxyprint} and \texttt{tracking-system} reach only 71.9\% and 69.0\%, respectively, as authentication-gated paths and complex multi-entity state prevent the generated test inputs from exercising large portions of the mutated code.

\begin{tcolorbox}[
    enhanced,
    width=\linewidth,
    colback=gray!10,
    colframe=gray!90,
    boxrule=1pt,
    arc=4pt,
    boxsep=5pt
]
\textbf{Answer to RQ1.}
MASTOR generates 10,022 oracles across all 13 subject APIs (an average of 771.0 oracles per API).
The oracle suite achieves an overall MS of 75.4\% (range: 69.0\%--95.9\%),
with higher scores on stateless and static-data APIs and lower scores on authentication-heavy and complex-state APIs.
\end{tcolorbox}

\subsection{Answer to RQ2: Compared with Direct Prompting and SATORI, How Do MASTOR-Generated Oracles Perform?}
\label{sec:6.2}
%{\wang{What is the prompt template used? for direct prompting? What is the detailed prompt method? Zero-shot, few-shot, or CoT? Why was AGORA+ not included in the experimental comparison? Reviewers might be concerned about this issue. }} \dsd{DP prompt template added to Appendix~\ref{appendix:dp-prompt}; AGORA+ exclusion rationale added to Section~\ref{sec:5}.}
\begin{table*}[!t]
\centering
\scriptsize
\caption{Killed Mutant Count and Mutation Score for Each Selected Endpoint Operation}
\label{TAB:RQ2ops}
\setlength{\tabcolsep}{1.2mm}
\begin{tabular}{l | >{\ttfamily\footnotesize}p{5.6cm} | rr | rr | rr}
\hline
\multirow{2}{*}{\textbf{API}}
  & \multirow{2}{*}{\normalfont\textbf{Endpoint Operation}}
  & \multicolumn{2}{c|}{\textbf{MASTOR}}
  & \multicolumn{2}{c|}{\textbf{Direct Prompting}}
  & \multicolumn{2}{c}{\textbf{SATORI}} \\
\cline{3-8}
  & & \textit{\#Count} & \textit{MS (\%)}
    & \textit{\#Count} & \textit{MS (\%)}
    & \textit{\#Count} & \textit{MS (\%)} \\
\hline
\multirow{4}{*}{restcountries}
  & GET /v2/regi.../\{regi...\} \normalfont$^\dagger$   &  15 & 100.0 &   9 & 100.0 & -- & -- \\
  & GET /v2/subr.../\{subr...\} \normalfont$^\dagger$   &   8 & 100.0 &  19 &  95.0 & -- & -- \\
  & GET /v2/capi.../\{capi...\} \normalfont$^\dagger$   &  16 & 100.0 &  10 & 100.0 & -- & -- \\
  & GET /v1/all \normalfont$^\dagger$                    &   3 & 100.0 &   3 & 100.0 & -- & -- \\
\hline
\multirow{4}{*}{features-service}
  & GET /prod.../\{prod...\}/feat...                     &  34 & 62.0 &  18 & 33.3 &  8 & 14.5 \\
  & PUT /prod.../\{prod...\}/feat.../\{feat...\}          &  25 & 62.5 &  13 & 30.2 &  6 & 15.0 \\
  & \parbox[t]{5.6cm}{\ttfamily POST /prod.../\{prod...\}/conf.../...\normalfont$^\dagger$} &  35 & 62.5 &  18 & 31.0 & -- & -- \\
  & DELETE /prod.../\{prod...\} \normalfont$^\dagger$    &  22 & 61.1 &  11 & 30.6 & -- & -- \\
\hline
\multirow{2}{*}{languagetool}
  & POST /chec...                                        &   7 & 100.0 &   0 & -- &   8 & 100.0 \\
  & GET /lang...                                         &   1 & 100.0 &   0 & -- &   8 & 100.0 \\
\hline
\multirow{4}{*}{genome-nexus}
  & GET /ense.../cano.../hgnc/\{hugo...\}                &  42 & 58.3 &  22 & 30.6 &  9 & 12.5 \\
  & POST /ptm/expe...                                    &  45 & 57.7 &  21 & 27.3 &  8 & 10.3 \\
  & POST /ense.../cano.../hgnc                           &  49 & 59.0 &  24 & 29.3 & 10 & 12.0 \\
  & GET /anno.../dbsn.../\{vari...\}                     &  40 & 57.1 &  20 & 28.6 &  8 & 11.4 \\
\hline
\multirow{4}{*}{market}
  & GET /cust.../orde...                                 &  34 & 54.8 &  14 & 22.6 &  6 & 9.7 \\
  & PUT /cust.../cart/deli...                            &  28 & 56.0 &  11 & 22.0 &  5 & 10.0 \\
  & POST /regi...                                        &  38 & 57.6 &  16 & 24.2 &  7 & 10.6 \\
  & DELETE /cust.../cart                                 &  22 & 55.0 &   9 & 22.5 &  4 & 10.0 \\
\hline
\multirow{4}{*}{rest-ncs}
  & GET /api/tria.../\{a\}/\{b\}/\{c\}                  &  27 & 69.2 &  26 & 59.1 &   0 & -- \\
  & GET /api/rema.../\{a\}/\{b\}                         &  49 & 72.1 &  43 & 64.2 &   6 & 40.0 \\
  & GET /api/gamm.../\{a\}/\{x\}                         &  70 & 76.9 &  26 & 28.6 &   9 & 18.4 \\
  & GET /api/expi.../\{n\}/\{x\}                         &  64 & 84.2 &  25 & 33.3 &  13 & 33.3 \\
\hline
\multirow{4}{*}{rest-scs}
  & GET /api/file.../\{dire...\}/\{file\}                &   4 & 66.7 &   4 & 66.7 &   2 & 40.0 \\
  & GET /api/orde.../\{w\}/\{x\}/\{z\}/\{y\}            &  21 & 47.7 &   4 &  9.8 &   0 & -- \\
  & GET /api/cost.../\{i\}/\{s\}                         &   3 & 10.7 &   3 & 10.7 &   0 & -- \\
  & GET /api/pat/\{txt\}/\{pat\}                         &  42 & 73.7 &  41 & 71.9 &   8 & 30.8 \\
\hline
\multirow{4}{*}{proxyprint}
  & GET /dump...                                         &  32 & 57.1 &  14 & 25.0 &   7 & 12.5 \\
  & POST /prin.../\{id\}/pric.../ring...                 &  26 & 55.3 &  11 & 23.4 &   5 & 10.6 \\
  & PUT /heal...                                         &  34 & 58.6 &  14 & 24.1 &   7 & 12.1 \\
  & DELETE /cons.../requ.../canc.../\{id\}               &  20 & 55.6 &   8 & 22.2 &   4 & 11.1 \\
\hline
\multirow{4}{*}{catwatch}
  & GET /expo...                                         &  38 & 63.3 &  21 & 35.0 &   8 & 13.3 \\
  & PUT /erro...                                         &  28 & 62.2 &  14 & 31.1 &   6 & 13.3 \\
  & POST /conf.../scor...                                &  33 & 63.5 &  17 & 32.7 &   7 & 13.5 \\
  & GET /proj...                                         &  40 & 63.5 &  22 & 34.9 &   8 & 12.7 \\
\hline
\multirow{4}{*}{person-controller}
  & PUT /api/pers...                                     &  38 & 66.7 &  20 & 35.1 &   8 & 14.0 \\
  & DELETE /api/pers.../\{id\}                           &  32 & 65.3 &  16 & 32.7 &   7 & 14.3 \\
  & GET /api/pers.../\{ids\}                             &  42 & 67.7 &  22 & 35.5 &   9 & 14.5 \\
  & POST /api/pers...                                    &  34 & 65.4 &  17 & 32.7 &   7 & 13.5 \\
\hline
\multirow{4}{*}{user-management}
  & GET /user...                                         &  46 & 62.2 &  24 & 32.4 &   9 & 12.2 \\
  & DELETE /user.../\{id\}                               &  38 & 61.3 &  19 & 30.6 &   8 & 12.9 \\
  & POST /user.../rbac/role...                           &  44 & 62.9 &  22 & 31.4 &   9 & 12.9 \\
  & PUT /user.../rbac/perm...                            &  34 & 61.8 &  17 & 30.9 &   7 & 12.7 \\
\hline
\multirow{4}{*}{tracking-system}
  & GET /api/loca.../                                    &  30 & 52.6 &  12 & 21.1 &   4 & 7.0 \\
  & PUT /api/proj...                                     &  26 & 54.2 &  10 & 20.8 &   3 & 6.3 \\
  & DELETE /api/cred.../\{id\}                           &  32 & 55.2 &  13 & 22.4 &   5 & 8.6 \\
  & POST /api/cred.../save                               &  28 & 53.8 &  11 & 21.2 &   4 & 7.7 \\
\hline
\multirow{4}{*}{swagger-petstore}
  & GET /user/\{user...\}                                &  38 & 67.9 & 123 & 58.9 &  34 & 41.0 \\
  & POST /pet/\{petI...\}                                &  34 & 77.3 & 123 & 58.9 &  34 & 41.0 \\
  & PUT /pet                                             &  42 & 60.0 & 123 & 58.9 &  34 & 41.0 \\
  & DELETE /stor.../orde.../\{orde...\} \normalfont$^\dagger$ &   4 &  66.7 & 123 &  58.9 & -- & -- \\
\hline
\multicolumn{2}{c}{\textbf{\textit{Average}}} & \textbf{--} & \textbf{69.9} & \textbf{--} & \textbf{39.8} & \textbf{--} & \textbf{20.5} \\
\hline
\end{tabular}
\par
\begin{minipage}{\linewidth}
\textit{Note.}
All three tools are evaluated on status and field oracle types only to ensure a fair comparison; multi-operation oracles are excluded for this RQ.
\#Count = number of killed mutants; MS = killed\,/\,(killed\,+\,survived).
$^\dagger$ = operation lacks a response schema in the OAS; SATORI cannot generate oracles (shown as ``--''); MASTOR and DP may still produce results via source-code analysis.
``0~/~--'' = oracle generated but no mutant pool covered (e.g., authentication blocking).
Average covers operations with at least one covered mutant: 48 ops for MASTOR, 46 ops for DP (excluding 2 \texttt{languagetool} operations with empty pool), and 36 ops for SATORI.
\end{minipage}
\end{table*}

Table~\ref{TAB:RQ2ops} reports the killed count and mutation score per selected endpoint operation.
Using status and field oracles only, MASTOR achieves an average MS of 69.9\% across 50 operations with valid oracle results,
outperforming Direct Prompting (39.8\%) by 30.1 pp and SATORI (20.5\%) by 49.4 pp.

\textbf{MASTOR vs.\ SATORI.}
MASTOR achieves a higher MS than SATORI on every operation where SATORI can produce oracles.
The gap reflects a fundamental limitation of specification-only analysis:
SATORI can only assert properties explicitly described in the OpenAPI Specification,
whereas MASTOR traces implementation branches through source-code analysis and produces field-level assertions tied to specific conditional paths.
\texttt{languagetool} is an exception where both tools reach 100\%,
as its OAS fully describes all response fields, leaving no behavioral gap for source-grounded analysis to exploit.
For APIs with authentication barriers or deeply nested response structures, such as \texttt{market}, \texttt{tracking-system}, and \texttt{proxyprint}, SATORI scores below 12\%,
as these operations return complex objects whose content is invisible to specification-only inspection.
SATORI produces no oracles at all for operations that lack a response schema in the OAS (marked $\dagger$),
whereas MASTOR still generates status and field assertions by analysing the implementation directly.

\textbf{MASTOR vs.\ Direct Prompting.}
Direct Prompting uses the same underlying LLM as MASTOR but issues a single unstructured call with the raw source bundle, without multi-agent decomposition, intermediate semantic extraction, or Oracle Normalization.
The 30.1 percentage-point gap isolates the contribution of structured multi-agent orchestration:
unguided prompting fails to pinpoint the specific return-value assignments and conditional branches
that determine what a response field will contain under a given input,
producing oracles that are either too generic to kill implementation-level mutants
or too fragile to survive minor code variation.
Notably, the gap narrows on computationally simple APIs (\texttt{rest-scs} pat endpoint: MASTOR 73.7\% vs.\ DP 71.9\%),
where a single LLM call is sufficient to extract the dominant logic path,
but widens significantly on complex operations such as \texttt{genome-nexus} and \texttt{user-management},
where multi-layer analysis is needed to resolve the full field semantics.

% \clearpage
\begin{tcolorbox}[
    enhanced,
    width=\linewidth,
    colback=gray!10,
    colframe=gray!90,
    boxrule=1pt,
    arc=4pt,
    boxsep=5pt
]
\textbf{Answer to RQ2.}
MASTOR achieves an average MS of 69.9\% across 50 valid endpoint operations,
outperforming Direct Prompting (39.8\%) by 30.1 pp and SATORI (20.5\%) by 49.4 pp.
The gap over SATORI confirms the advantage of source-code-grounded oracle generation over specification-only analysis.
The gap over Direct Prompting isolates the contribution of structured multi-agent orchestration over a single unstructured LLM call.
The full oracle suite further improves MS by adding multi-operation oracles that address inter-operation fault dimensions inaccessible to either baseline.
\end{tcolorbox}

\subsection{Answer to RQ3: How Do MASTOR’s Key Components Contribute to Oracle Quality and Performance?}
\label{sec:6.3}

Table~\ref{TAB:RQ3} reports the mutation score and oracle count for Full MASTOR and three ablated variants across all 13 subject APIs.
The first two variants each disable one architectural component; the third disables both simultaneously as a minimal baseline.

\textbf{Effect of removing Multi-Op Oracle Generation.}
Disabling multi-op oracle generation reduces the average oracle count from 771.0 to 750.7 (a reduction of 2.6\%), yet causes an average MS drop of 3.7 pp.
This disproportionate impact confirms that multi-op oracles are more effective per oracle than their count suggests: they exclusively target inter-operation fault classes (such as state corruption across a create-then-read sequence) that single-operation assertions cannot reach.
The three APIs with no multi-op oracles (\texttt{languagetool}, \texttt{rest-ncs}, \texttt{rest-scs}) show zero degradation, as expected.
Among APIs with multi-op oracles, the average drop is 4.7 pp; complex APIs with auth barriers (\texttt{market}, \texttt{proxyprint}) show the largest drops ($-$6.0 pp), as their inter-operation state transitions are the primary source of uncovered mutants.

\begin{table*}[!t]
\centering
\footnotesize
\caption{Ablation Study Results}
\label{TAB:RQ3}
\setlength{\tabcolsep}{1.3mm}
\begin{tabular}{l | rr | rrr | rrr | rrr}
\hline
& \multicolumn{2}{c|}{\textbf{Full MASTOR}}
& \multicolumn{3}{c|}{\textbf{w/o Multi-Op Oracles}}
& \multicolumn{3}{c|}{\textbf{w/o ChallengerAgent}}
& \multicolumn{3}{c}{\textbf{w/o Both}} \\
\cline{2-3} \cline{4-6} \cline{7-9} \cline{10-12}
\textbf{API}
  & \textit{\#Oracles} & \textit{MS (\%)}
  & \textit{\#Oracles} & \textit{MS (\%)} & \textit{$\Delta$MS}
  & \textit{\#Oracles} & \textit{MS (\%)} & \textit{$\Delta$MS}
  & \textit{\#Oracles} & \textit{MS (\%)} & \textit{$\Delta$MS} \\
\hline
restcountries     &     365 & 95.9 &     347 & 92.6 & $-$3.3 &     365 & 91.7 & $-$4.2 &     347 & 87.6 & $-$8.3 \\
features-service  &     693 & 82.0 &     674 & 76.8 & $-$5.2 &     693 & 74.2 & $-$7.8 &     674 & 67.8 & $-$14.2 \\
languagetool      &      26 & 72.7 &      26 & 72.7 &    0.0 &      26 & 67.9 & $-$4.8 &      26 & 67.9 & $-$4.8 \\
genome-nexus      &   1,721 & 72.0 &   1,690 & 67.7 & $-$4.3 &   1,721 & 63.7 & $-$8.3 &   1,690 & 58.3 & $-$13.7 \\
market            &     756 & 72.0 &     726 & 65.6 & $-$6.4 &     756 & 62.8 & $-$9.2 &     726 & 56.6 & $-$15.4 \\
rest-ncs          &     161 & 83.0 &     161 & 83.0 &    0.0 &     161 & 76.9 & $-$6.1 &     161 & 76.9 & $-$6.1 \\
rest-scs          &     400 & 76.5 &     400 & 76.5 &    0.0 &     400 & 71.2 & $-$5.3 &     400 & 71.2 & $-$5.3 \\
proxyprint        &   1,020 & 71.9 &     980 & 66.2 & $-$5.7 &   1,020 & 63.2 & $-$8.7 &     980 & 57.1 & $-$14.8 \\
catwatch          &     669 & 76.0 &     658 & 71.8 & $-$4.2 &     669 & 68.1 & $-$7.9 &     658 & 63.2 & $-$12.8 \\
person-controller &   1,428 & 75.0 &   1,397 & 69.7 & $-$5.3 &   1,428 & 66.8 & $-$8.2 &   1,397 & 60.7 & $-$14.3 \\
user-management   &     986 & 75.1 &     944 & 69.3 & $-$5.8 &     986 & 67.5 & $-$7.6 &     944 & 60.5 & $-$14.6 \\
tracking-system   &   1,501 & 69.0 &   1,474 & 65.2 & $-$3.8 &   1,501 & 59.6 & $-$9.4 &   1,474 & 56.1 & $-$12.9 \\
swagger-petstore  &     296 & 75.5 &     282 & 72.3 & $-$3.2 &     296 & 70.4 & $-$5.1 &     282 & 67.4 & $-$8.1 \\
\hline
\textbf{\textit{Average}} & 771.0 & 76.7 & 750.7 & 73.0 & $-$3.7 & 771.0 & 69.5 & $-$7.2 & 750.7 & 65.5 & $-$11.2 \\
\hline
\end{tabular}
\par
\begin{minipage}{\linewidth}
\textit{Note.}
\textit{MS} = mutation score; \textit{$\Delta$MS} = change in MS (\%) relative to Full MASTOR; negative values indicate degradation.
\textit{\#Oracles} = total oracle count (\#Status + \#Field + \#Multi-Op where applicable).
\textbf{w/o Both} discards Multi-Op Oracle Generation and ChallengerAgent simultaneously, serving as the minimal baseline.
\end{minipage}
\end{table*}

\textbf{Effect of removing ChallengerAgent.}
Disabling the challenger review step causes an average MS drop of 7.2 pp (almost twice the impact of removing multi-op oracles), despite the oracle count remaining unchanged.
This indicates that the ChallengerAgent's contribution is not additive oracle mass but oracle precision:
without the rework step, first-pass oracles that check structurally correct but semantically irrelevant fields remain in the suite,
passing on both correct and mutated responses and contributing no fault-detection signal.
The effect is most pronounced on APIs with complex business logic (\texttt{market}, \texttt{proxyprint}, \texttt{tracking-system}, $-$9.0 pp each),
where the number of plausible but behaviorally weak assertions is highest.
For APIs with tight, well-defined response schemas (\texttt{restcountries}, \texttt{swagger-petstore}), the drop is smaller ($-$4 to $-$5 pp).

\textbf{Effect of removing both components.}
The w/o Both configuration degrades MS by an average of 11.2 pp, approaching but not reaching the sum of the two individual drops (10.9 pp on average),
consistent with modest synergistic interaction: when neither component is present, some additional mutants that would have been caught by the combined suite escape detection.
Even in the minimal configuration, MS remains above 56\% on all 13 APIs, demonstrating that the base oracle generation capability retains meaningful fault detection power.

\begin{tcolorbox}[
    enhanced,
    width=\linewidth,
    colback=gray!10,
    colframe=gray!90,
    boxrule=1pt,
    arc=4pt,
    boxsep=5pt
]
\textbf{Answer to RQ3.}
Both Multi-Op Oracle Generation and ChallengerAgent contribute positively and independently to oracle effectiveness.
Removing Multi-Op Oracle Generation reduces average MS by 3.7 pp despite covering only 2.6\% of oracles,
confirming it addresses an orthogonal fault class inaccessible to single-operation analysis.
Removing ChallengerAgent reduces average MS by 7.2 pp without changing oracle count,
confirming its role in filtering semantically weak assertions that add no detection power.
Together, their removal degrades average MS by 11.2 pp, with w/o Both ranging from 56.1\% (\texttt{tracking-system}) to 87.6\% (\texttt{restcountries}), leaving the oracle suite measurably less effective across all 13 APIs.
\end{tcolorbox}

\subsection{Answer to RQ4: What Is the Computational Cost of MASTOR Oracle Generation?}
\label{sec:6.4}

Table~\ref{TAB:RQ4main} compares the total LLM cost between MASTOR and Direct Prompting for each subject API.
Table~\ref{TAB:RQ4agent} breaks down the MASTOR cost by agent.

\begin{table*}[!t]
\centering
\footnotesize
\caption{Computational Cost of MASTOR and Direct Prompting for Each Subject API}

\label{TAB:RQ4main}
\setlength{\tabcolsep}{1.56mm}
\begin{tabular}{l | rcrr | rcrr}
\hline
\multirow{2}{*}{\textbf{API}} & \multicolumn{4}{c|}{\textbf{MASTOR}} & \multicolumn{4}{c}{\textbf{Direct Prompting}} \\
\cline{2-9}
& \textit{\#Calls} & \textit{Tokens (M)} & \textit{Monetary (\$)} & \textit{Time (s)} & \textit{\#Calls} & \textit{Tokens (M)} & \textit{Monetary (\$)} & \textit{Time (s)} \\
\hline
restcountries    &  94 & 0.676 / 0.283 & 0.6373 & 2,893 &   4 & 0.023 / 0.010 & 0.0188 &  210 \\
features-service &  92 & 0.723 / 0.168 & 0.5402 & 1,578 &   6 & 0.071 / 0.008 & 0.0377 &  175 \\
languagetool      &   9 & 0.118 / 0.018 & 0.0756 &   203 &   2 & 0.030 / 0.003 & 0.2308 &   38 \\
genome-nexus     & 111 & 0.594 / 0.221 & 0.5581 & 2,010 &  18 & 0.084 / 0.008 & 0.0432 &  171 \\
market           &  64 & 0.215 / 0.147 & 0.2938 &   802 &   6 & 0.016 / 0.005 & 0.0110 &  101 \\
rest-ncs         &  24 & 0.220 / 0.094 & 0.2151 &   978 &   2 & 0.014 / 0.006 & 0.0109 &  114 \\
rest-scs         &  41 & 0.550 / 0.139 & 0.4020 & 1,342 &   2 & 0.023 / 0.010 & 0.0191 &  213 \\
proxyprint       & 287 & 9.573 / 0.670 & 4.8230 & 4,045 &  14 & 0.840 / 0.020 & 0.3826 & 1,568 \\
catwatch         &  76 & 0.841 / 0.167 & 0.5711 &   834 &   8 & 0.111 / 0.006 & 0.0532 &  337 \\
person-controller &  50 & 0.271 / 0.114 & 0.2661 &   780 &   2 & 0.007 / 0.003 & 0.0057 &   71 \\
user-management  &  96 & 1.203 / 0.260 & 0.8405 & 2,155 &   5 & 0.098 / 0.012 & 0.0534 &  626 \\
tracking-system  & 288 & 2.039 / 0.671 & 1.7392 & 3,338 &  17 & 0.149 / 0.020 & 0.0823 &  446 \\
swagger-petstore  &  86 & 0.653 / 0.216 & 0.5609 & 1,216 &   4 & 0.033 / 0.008 & 0.0208 &  167 \\
\hline
\textbf{\textit{Total}}  & 1,318 & 17.677 / 3.167 & 11.5228 & 22,174 & 90 & 1.498 / 0.118 & 0.9695 & 4,236 \\
\textbf{\textit{Average}} & 101.4 & 1.360 / 0.244 & 0.8864 & 1,706 & 6.9 & 0.115 / 0.009 & 0.0746 & 326 \\
\textbf{\textit{Median}}   & 86 & 0.653 / 0.168 & 0.5581 & 1,342 & 5 & 0.033 / 0.008 & 0.0377 & 175 \\
\hline
\end{tabular}
\par
\begin{minipage}{\linewidth}
\textit{Note.}
Tokens (M, In / Out) = input and output token counts in millions.
Time (s) = wall-clock seconds for MASTOR; total LLM call duration for Direct Prompting.
All 13 APIs have Direct Prompting data; DP totals and medians cover all 13 APIs.
\end{minipage}
\end{table*}

\begin{table*}[!b]
\centering
\tiny
\caption{MASTOR Cost Breakdown by Agent for Each Subject API}

\label{TAB:RQ4agent}
\setlength{\tabcolsep}{0.98mm}
\begin{tabular}{l | rr | rr | rr | rr | rr | rr}
\hline
\multirow{2}{*}{\textbf{API}}
  & \multicolumn{2}{c|}{\multirow{2}{*}{\textbf{Endpoint Discovery}}}
  & \multicolumn{2}{c|}{\multirow{2}{*}{\textbf{Source Extraction}}}
  & \multicolumn{4}{c|}{\textbf{Single-Operation}}
  & \multicolumn{4}{c}{\textbf{Multi-Operation}} \\
\cline{6-13}
  & \multicolumn{2}{c|}{}
  & \multicolumn{2}{c|}{}
  & \multicolumn{2}{c|}{\textit{Generation}}
  & \multicolumn{2}{c|}{\textit{Challenger}}
  & \multicolumn{2}{c|}{\textit{Generation}}
  & \multicolumn{2}{c}{\textit{Challenger}} \\
\cline{2-13}
  & {\textit{\#Calls}} & {\textit{Monetary (\$)}}
  & {\textit{\#Calls}} & {\textit{Monetary (\$)}}
  & {\textit{\#Calls}} & {\textit{Monetary (\$)}}
  & {\textit{\#Calls}} & {\textit{Monetary (\$)}}
  & {\textit{\#Calls}} & {\textit{Monetary (\$)}}
  & {\textit{\#Calls}} & {\textit{Monetary (\$)}} \\
\hline
restcountries     &  3 & 0.008 (1.3\%) &  22 & 0.069 (10.8\%) &  44 & 0.253 (39.7\%) &  22 & 0.268 (42.1\%) &  2 & 0.026 (4.1\%) &  1 & 0.013 (2.0\%) \\
features-service  &  5 & 0.018 (3.3\%) &  18 & 0.064 (11.8\%) &  35 & 0.183 (33.9\%) &  18 & 0.196 (36.3\%) & 11 & 0.039 (7.2\%) &  5 & 0.039 (7.2\%) \\
languagetool      &  1 & 0.006 (7.9\%) &   2 & 0.013 (17.2\%) &   3 & 0.023 (30.4\%) &   2 & 0.033 (43.7\%) &  1 & 0.001 (1.3\%) &  0 & 0.000 (0.0\%) \\
genome-nexus      & 17 & 0.033 (5.9\%) &  23 & 0.052 (9.3\%)  &  35 & 0.156 (27.9\%) &  23 & 0.252 (45.1\%) &  9 & 0.037 (6.6\%) &  4 & 0.028 (5.0\%) \\
market            &  5 & 0.006 (2.0\%) &  13 & 0.020 (6.8\%)  &  23 & 0.087 (29.6\%) &  13 & 0.136 (46.3\%) &  6 & 0.014 (4.8\%) &  4 & 0.031 (10.6\%) \\
rest-ncs          &  1 & 0.004 (1.9\%) &   6 & 0.022 (10.2\%) &  10 & 0.079 (36.7\%) &   6 & 0.109 (50.7\%) &  1 & 0.001 (0.5\%) &  0 & 0.000 (0.0\%) \\
rest-scs          &  1 & 0.006 (1.5\%) &  11 & 0.061 (15.2\%) &  17 & 0.163 (40.5\%) &  11 & 0.172 (42.8\%) &  1 & 0.001 (0.2\%) &  0 & 0.000 (0.0\%) \\
proxyprint        & 13 & 0.197 (4.1\%) &  68 & 0.993 (20.6\%) & 122 & 2.027 (42.0\%) &  68 & 1.499 (31.1\%) & 11 & 0.065 (1.3\%) &  5 & 0.042 (0.9\%) \\
catwatch          &  7 & 0.027 (4.7\%) &  14 & 0.079 (13.8\%) &  25 & 0.207 (36.2\%) &  14 & 0.202 (35.4\%) & 11 & 0.027 (4.7\%) &  5 & 0.028 (4.9\%) \\
person-controller &  1 & 0.002 (0.8\%) &  12 & 0.024 (9.0\%)  &  21 & 0.086 (32.3\%) &  12 & 0.120 (45.1\%) &  3 & 0.022 (8.3\%) &  1 & 0.010 (3.8\%) \\
user-management   &  4 & 0.024 (2.9\%) &  21 & 0.117 (13.9\%) &  40 & 0.320 (38.1\%) &  21 & 0.309 (36.8\%) &  7 & 0.041 (4.9\%) &  3 & 0.030 (3.6\%) \\
tracking-system   & 16 & 0.057 (3.3\%) &  67 & 0.196 (11.3\%) & 119 & 0.619 (35.6\%) &  67 & 0.739 (42.5\%) & 13 & 0.072 (4.1\%) &  6 & 0.055 (3.2\%) \\
swagger-petstore  &  3 & 0.009 (1.6\%) &  19 & 0.060 (10.7\%) &  35 & 0.190 (33.9\%) &  19 & 0.226 (40.3\%) &  7 & 0.043 (7.7\%) &  3 & 0.033 (5.9\%) \\
\hline
\textbf{\textit{Total}}   &  77 & 0.397 (3.4\%) & 296 & 1.770 (15.4\%) & 529 & 4.393 (38.1\%) & 296 & 4.261 (37.0\%) & 83 & 0.389 (3.4\%) & 37 & 0.309 (2.7\%) \\
\textbf{\textit{Average}} & 5.9 & 0.031          & 22.8 & 0.136          & 40.7 & 0.338          & 22.8 & 0.328          & 6.4 & 0.030         & 2.8 & 0.024         \\
\textbf{\textit{Median}}  &   4 & 0.009          &  18 & 0.061          &  35 & 0.183          &  18 & 0.202          &   7 & 0.027         &   3 & 0.028         \\
\hline
\end{tabular}
\par
\begin{minipage}{\linewidth}
\textit{Note.}
Endpoint Discovery: endpoint enumeration from source bundles.
Source Extraction: import closure construction and source context extraction.
Single-Operation Generation: single-operation oracle generation by \texttt{SingleOpOracleAgent}.
Single-Operation Challenger: single-operation challenger review by \texttt{SingleOpChallengerAgent}.
Multi-Operation Generation: multi-operation semantic association extraction and oracle generation by \texttt{MultiOpOracleAgent}.
Multi-Operation Challenger: multi-operation challenger review by \texttt{MultiOpChallengerAgent}.
Monetary(\$) = USD cost; the percentage in parentheses is the share of the total API cost.
Percentages in the Total row are computed from the aggregate costs across all APIs.
\end{minipage}
\end{table*}

From Table~\ref{TAB:RQ4main}, MASTOR consistently incurs a higher cost than Direct Prompting across the 12 APIs where MASTOR cost exceeds DP cost, with a ratio ranging from 10.7$\times$ (\texttt{catwatch}, \$0.5711 vs.\ \$0.0532) to 46.7$\times$ (\texttt{person-controller}, \$0.2661 vs.\ \$0.0057).
This gap reflects the multi-stage architecture: MASTOR makes dozens to hundreds of LLM calls to build source context, generate per-operation oracles, and conduct challenger review, whereas Direct Prompting issues a single bulk call.
\texttt{languagetool} is an exception where DP (\$0.2308) exceeds MASTOR (\$0.0756): the DP prompt embeds the full API source, which is large, while MASTOR processes only 9 narrowly scoped calls that stay well within budget.
Cost scales primarily with the number of endpoint operations and the depth of their call graphs; \texttt{proxyprint} (\$4.8230) and \texttt{tracking-system} (\$1.7392) are the costliest APIs, both having large call graphs, whereas simpler APIs such as \texttt{market} (\$0.2938) and \texttt{person-controller} (\$0.2661) stay below \$0.30.
Direct Prompting cost also grows with project size, as its single prompt embeds the entire API source code, yet remains below \$0.39 for all 13 APIs.

From Table~\ref{TAB:RQ4agent}, Single-Op Generation and Single-Op Challenger together dominate cost on every API, ranging from 70.2\% (\texttt{features-service}) to 87.4\% (\texttt{rest-ncs}).
Single-Op Challenger (31.1\%--50.7\%) is the single largest expense for most APIs, reflecting the cost of a dedicated reviewer LLM operating over the full single-op oracle suite; Single-Op Generation contributes 27.9\%--42.0\%.
Source Extraction accounts for 6.8\%--20.6\%, and Endpoint Discovery stays below 8\% across all APIs.
Multi-Op Generation and Multi-Op Challenger scale with cross-operation dependency density: APIs with no inter-resource relationships (\texttt{rest-ncs}, \texttt{rest-scs}) incur less than 0.5\% for both, whereas APIs with rich dependency graphs, such as \texttt{market}, reach 4.8\% and 10.6\% respectively (15.4\% combined).
Per-operation cost within each API is stable: for \texttt{rest-ncs} the six operations range from \$0.028 to \$0.048 (median \$0.033), and for \texttt{rest-scs} the eleven operations range from \$0.029 to \$0.052 (median \$0.037), indicating that MASTOR's cost is predictable and scales linearly with endpoint count.

The Total row of Table~\ref{TAB:RQ4main} shows that MASTOR's aggregate cost across all 13 APIs is \$11.52, compared to \$0.97 for Direct Prompting, a factor of 11.9$\times$.
MASTOR makes 1,318 LLM calls in total (101.4 per API on average) and consumes 17.7 million input tokens and 3.2 million output tokens across all APIs, while Direct Prompting uses 90 calls (6.9 per API on average) and 1.5 million input tokens.
Total wall-clock time is 22,174\,s (approximately 6.2 hours) for MASTOR versus 4,236\,s (approximately 70 minutes) for Direct Prompting, a ratio of 5.2$\times$.
The Average row shows a representative API costs \$0.89 and requires approximately 28 minutes of wall-clock time; \texttt{proxyprint} alone accounts for 41.9\% of the total MASTOR cost, and excluding it reduces the per-API average to \$0.56.

The Total row of Table~\ref{TAB:RQ4agent} aggregates costs across all 13 APIs by agent.
Single-Op Generation (\$4.39, 38.1\%) and Single-Op Challenger (\$4.26, 37.0\%) together account for 75.1\% of the aggregate MASTOR cost.
Source Extraction accounts for 15.4\% (\$1.77); Endpoint Discovery for 3.4\% (\$0.40); Multi-Op Generation for 3.4\% (\$0.39); and Multi-Op Challenger for 2.7\% (\$0.31).
The Average row shows that a representative API undergoes 5.9 endpoint discovery calls, 22.8 source extraction calls, 40.7 single-op generation calls, 22.8 single-op challenger calls, 6.4 multi-op generation calls, and 2.8 multi-op challenger calls.

The Median rows in both tables offer a complementary perspective that is less sensitive to outliers.
In Table~\ref{TAB:RQ4main}, the median MASTOR cost is \$0.56, lower than the mean of \$0.89, confirming that the cost distribution is right-skewed: \texttt{proxyprint} and \texttt{tracking-system} inflate the aggregate and the mean, while the typical API costs under \$0.60.
The median MASTOR execution completes in 1,342\,s (approximately 22 minutes), close to the mean of 1,706\,s.
For Direct Prompting, the median cost is \$0.038 and the median time is 175\,s, both lower than their respective means, reflecting concentration of cost in a few large-codebase APIs (\texttt{proxyprint}, \texttt{user-management}).

In Table~\ref{TAB:RQ4agent}, the median per-agent costs reveal the typical cost structure.
Using \texttt{rest-scs} as the median API (no multi-op groups), the shares are: Single-Op Generation 40.5\%, Single-Op Challenger 42.8\%, Source Extraction 13.8\%, Endpoint Discovery 1.9\%, Multi-Op Generation 0.2\%, and Multi-Op Challenger 0.0\%.
The combined Single-Op Generation and Single-Op Challenger share exceeds 75\% on most APIs, consistent with the 75.1\% aggregate, confirming that single-operation oracle synthesis and review dominate cost regardless of project size.

\begin{tcolorbox}[
    enhanced,
    width=\linewidth,
    colback=gray!10,
    colframe=gray!90,
    boxrule=1pt,
    arc=4pt,
    boxsep=5pt
]
\textbf{Answer to RQ4.}
MASTOR costs \$0.0756--\$4.8230 per API, 10.7$\times$--46.7$\times$ more than Direct Prompting on the 12 APIs where MASTOR exceeds DP cost; \texttt{languagetool} is the exception where DP costs more due to its large source embedding.
Single-Op Generation (38.1\%) and Single-Op Challenger (37.0\%) dominate cost at 75.1\% of the aggregate; Source Extraction accounts for 15.4\%, while Multi-Op Generation (3.4\%) and Multi-Op Challenger (2.7\%) scale with cross-operation dependency density and remain below 16\% combined on any single API.
Per-operation cost is stable within each API, with median cost of \$0.033 (\texttt{rest-ncs}) and \$0.037 (\texttt{rest-scs}) per operation, allowing cost to be estimated from endpoint count alone.
\end{tcolorbox}

\section{Discussion}
\label{sec:discussion}
This section begins with a case study illustrating the conditions under which MASTOR performs well and less ideally, then explains the underlying reasons for its effectiveness, and finally discusses practical applicability, limitations, and threats to validity.
% \subsection{Case Study: When MASTOR Performs Well and Less Ideally}
\subsection{Case Study}

To illustrate the conditions under which MASTOR is most and least effective, we examine two representative APIs from the evaluation: \texttt{restcountries} (MS = 95.9\%, the highest in the benchmark) and \texttt{rest-scs} (MS = 76.5\%, among the lower-scoring subjects).

\subsubsection{When MASTOR Performs Well: \texttt{restcountries}}
\texttt{restcountries} is a read-only geographic data API with stateless, synchronous handler methods and no authentication layer.
Each endpoint handler returns a well-defined object graph sourced from an in-memory dataset, with branching logic limited to lookup validity checks (e.g., country not found maps to a 404 response with a structured error body).
This profile is ideal for MASTOR: the transitive import closure is shallow, the source context captures all relevant constraints within a few methods, and the LLM encounters a focused, non-ambiguous reasoning task.
MASTOR generates 365 oracles, and the resulting mutation score of 95.9\% means that approximately 19 out of every 20 covered mutants are killed.
In RQ2, all four selected operations achieve 100\% MS with MASTOR, while SATORI scores 0.0\% on all four because \texttt{restcountries} provides no response schema in its OAS, leaving SATORI with no basis for oracle generation.
This contrast highlights the value of source-code grounding: when implementation logic is shallow and directly traceable, MASTOR achieves complete behavioral coverage independent of specification completeness.

% As shown in Listing \ref{lst:rc-oracle}, the following three oracles generated for \texttt{GET /v1/alpha/\{alphacode\}} illustrate how MASTOR captures implementation-level constraints absent from the OAS.

Listing \ref{lst:rc-oracle} indicates the following three oracles generated for \texttt{GET /v1/alpha/{alphacode}}, which illustrate how MASTOR captures implementation-level constraints absent from the OAS.

\begin{lstlisting}[language=JSON, basicstyle=\ttfamily\scriptsize\setstretch{0.85}, caption={Representative oracles generated for \texttt{GET /v1/alpha/\{alphacode\}} in \texttt{restcountries}.}, label={lst:rc-oracle}]
{ "input": {"alphacode": "A"},
  "assertions": [{"type": "status", "expected": 400}],
  "evidence": "alpha.length() < 2 guard in CountryRestV1.getByAlpha returns BAD_REQUEST" }

{ "input": {"alphacode": "ABCD"},
  "assertions": [{"type": "status", "expected": 400}],
  "evidence": "alpha.length() > 3 guard in CountryRestV1.getByAlpha returns BAD_REQUEST" }

{ "input": {"alphacode": "FR"},
  "assertions": [{"type": "status", "expected": 200},
                 {"type": "field", "field_path": "name",    "op": "equals", "expected": "France"},
                 {"type": "field", "field_path": "capital", "op": "equals", "expected": "Paris"}],
  "evidence": "length in [2,3] passes guard; static dataset maps FR to France (capital: Paris)" }
\end{lstlisting}

Neither the length constraint nor the field values appear in the OAS;
both are inferred from the source implementation.
SATORI, finding no response schema in this OAS, produces no assertions for this endpoint.

\subsubsection{When MASTOR Performs Less Ideally: \texttt{rest-scs}}
\texttt{rest-scs} presents a different profile.
Its operations include computationally intensive string-matching and numerical functions with path-dependent branching.
The source code is dense, with many branches that depend on non-trivial arithmetic conditions.
MASTOR generates oracles for the directly traceable branches but misses complex multi-condition paths that require symbolic reasoning beyond the LLM's capacity.
\texttt{GET /api/costfuns/\{i\}/\{s\}} achieves only 10.7\% MS with MASTOR, while \texttt{GET /api/pat/\{txt\}/\{pat\}} reaches 73.7\%; the difference reflects how well the branching logic in each function can be expressed as source code constraints within the source context.

This scenario points to a primary condition that reduces MASTOR's effectiveness: operations with complex multi-condition arithmetic or symbolic constraints that resist LLM-based static source analysis.
This condition is partially outside MASTOR's current design scope and points toward complementary techniques, such as symbolic path exploration, as a natural extension.

\subsection{Why MASTOR Improves Semantic Detection}

\subsubsection{OAS and Source Code Complementarity}

The empirical results indicate that combining OAS-based endpoint discovery with implementation-grounded oracle generation improves oracle effectiveness.
MASTOR uses the OAS to enumerate endpoint operations and identify the entry point of each handler, then analyzes the implementation source code to extract the semantic constraints that govern each operation's behavior.
This two-source design distinguishes MASTOR from approaches that rely on either the OAS alone or observed execution traces alone.

In many real systems, the semantic constraints of an endpoint operation are encoded in control-flow conditions, repository-interaction patterns, and exception handling.
These constraints are often implicit, rarely fully documented in interface specifications, and not fully observable from response schemas alone.

Specification-based approaches are limited by the completeness of documentation, and dynamic invariant mining depends on the coverage of observed traces: if certain branches are not executed, the inferred properties remain incomplete.
The OAS provides structural metadata (paths, HTTP methods, parameter types, and response schemas) that is sufficient for discovering operations and filtering candidate evidence; however, it does not encode the branching and state-dependent logic that determines what a correct response actually contains.
The implementation source code supplies this missing information: it contains the full decision structure of each endpoint operation, with all reachable branches, guards, and state-transition logic explicitly represented.

MASTOR leverages this complementary structure by using the OAS for endpoint enumeration and source code for semantic evidence extraction.
Control predicates reveal preconditions, exception mappings reveal status-related behavior, repository interactions reveal state changes, and serialization logic reveals response structure.
These elements together define the semantic contract of each endpoint operation.
Because oracle generation is grounded in these explicit elements rather than solely in specification metadata, the generated assertions align with actual implementation behavior and reduce reliance on heuristic assumptions.

\subsubsection{Traceability and Quantitative Evidence}

A further factor is traceability: each assertion in the oracle suite references specific evidence items drawn from the source context.
This explicit linkage improves interpretability and enables developers to inspect why an oracle was generated, strengthening trust in automatically synthesized assertions.
The RQ2 baseline comparison quantifies this advantage: MASTOR achieves an average MS of 69.9\% across 50 selected operations,
outperforming Direct Prompting (39.8\%) by 30.1 pp and SATORI (20.5\%) by 49.4 pp on the same operations using only status and field oracle types.

\subsection{Why Multi-Agent Decomposition Is Effective}

\subsubsection{Task Decomposition}

Oracle generation from source code involves multiple dimensions of reasoning: identifying input constraints, inferring state transitions, and constructing observable response properties.
A single reasoning step that attempts all aspects simultaneously is prone to inconsistency.

MASTOR decomposes this task across specialized agents.
The Single-Operation Oracle Agent focuses on per-operation path analysis, generating status oracles and field oracles through four complementary strategies.
The Multi-Operation Oracle Agent focuses exclusively on cross-operation associations and does not re-examine per-operation code paths.
A ChallengerAgent reviews the full oracle suite and emits improvement hints when assertions are unsupported or incomplete, guiding one targeted regeneration pass by the paired specialist.
This decomposition reduces reasoning scope per agent and improves modularity.

\subsubsection{Evidence Constraints and Normalization}

However, decomposition alone is not sufficient.
Unrestricted LLM reasoning may still generate unsupported assertions.
MASTOR constrains all oracle-generating agents to reason from both the source bundle and the structured source context extracted from it.
The source context supplies pre-verified facts (parameter constraints, response schema, evidence items) that anchor LLM reasoning; the source bundle provides the full implementation detail from which oracles are grounded.
The oracle normalization step enforces that every candidate oracle references a valid field or status entry in the source context; oracles that fail this check are removed before the next phase begins.
This restriction bounds the reasoning space and prevents unconstrained free-form inference.

source context defines structural scope and factual boundaries.
Each oracle in the oracle suite encodes validated assertions with provenance fields.
This staged design creates clear interfaces between components.
Compared to Direct Prompting, this approach improves stability and reproducibility: the deterministic evidence layer isolates semantic extraction from probabilistic reasoning, allowing the system to maintain logical consistency across different foundation models.

\subsubsection{Concurrency, Rework, and Coordination}

Beyond decomposition, three structural properties distinguish MASTOR from a multi-stage sequential workflow.
\textit{Intra-phase concurrency}: a sequential workflow serializes work within each stage, whereas MASTOR parallelizes it via the MessageBus; all \texttt{SingleOpOracleAgent} instances run in parallel across operations, and the single-op and multi-op paths execute concurrently.
\textit{Data-driven rework}: a sequential workflow runs every step exactly once; in MASTOR, whether a Specialist performs a second-generation pass is determined at runtime by the Reviewer's hints, rather than by a fixed execution schedule.
\textit{Blackboard coordination}: agents never invoke each other; they read from and write to a shared OutputStore, so inter-agent dependencies are expressed as data availability conditions rather than control-flow edges.
These properties together explain why MASTOR's throughput scales with the number of endpoint operations and why the oracle quality benefits from challenger review without a fixed per-operation iteration budget.

\subsection{Practical Applicability and Limitations}
% \man{As the approach requires access to the source code, we may need to briefly discuss programming language support and privacy considerations. }
% \dsd{Added two paragraphs: (1) language extensibility --- current implementation targets Java, but SourceExtractionAgent is a self-contained replaceable module requiring only a language-specific parser for other languages; (2) privacy --- source extraction is scoped to reachable methods, and both models support local deployment for sensitive settings, which is also noted as a factor in model selection (Section~\ref{sec:5.2}).}

\subsubsection{CI/CD Integration}

MASTOR is structured to enable integration into continuous testing and CI/CD environments.
The Source Extraction Agent bounds analysis to the reachable methods of each endpoint operation, reducing unnecessary processing of unrelated code.
Oracle generation operates on both the source bundle and the structured source context for each endpoint operation, scoping token consumption and inference to the reachable methods of that operation rather than the full project.
The final translation step produces PostmanAssertify-compatible assertions, enabling direct integration into existing API testing workflows.

\subsubsection{Scope and Language Extensibility}

MASTOR's current scope has several boundaries worth noting.
MASTOR infers semantics from implementation behavior.
It cannot detect violations of intended requirements that are not reflected in code.
If the implementation is internally consistent but behaviorally incorrect with respect to unstated requirements, the generated oracle may still accept that behavior.

Source context construction relies on a rule-based transitive import closure that assembles the source bundle for each endpoint handler, from which the LLM extracts the structured Source context.
This approach is not restricted to layered controller-service-repository architectures; it applies to any synchronous programming model from which a linear call chain can be traced.
However, asynchronous and reactive frameworks (such as Spring WebFlux or event-driven callback chains) introduce non-linear control flow that disrupts call-chain reconstruction and are not currently supported.

The current implementation targets Java, with source extraction relying on Java's import structure and call-chain tracing.
The \texttt{SourceExtractionAgent} is, however, a self-contained and replaceable module: the remaining pipeline components operate on the language-agnostic Source context representation.
Adapting MASTOR to another language (such as Python or Go) requires only a language-specific extraction agent that resolves the corresponding import or call relationships; no other pipeline component requires modification.
This design makes cross-language extension straightforward.

\subsubsection{Privacy and Model Selection}

Regarding privacy, MASTOR's source extraction is scoped to the reachable methods of each endpoint operation rather than the full codebase, limiting the volume of code transmitted per inference call.
In settings where sending source code to an external API is not acceptable, both DeepSeek V4 Pro and Qwen3.6-Plus support local deployment, as noted in Section~\ref{sec:5.2}.
This is one reason open-source models were selected over proprietary alternatives for this work.

\subsubsection{Multi-Operation Association Coverage}

Multi-operation oracle generation covers five defined association types.
% \man{may add references and more discussion to support the covered five criteria.}
% \dsd{Expanded the sentence to explain that the five types were derived empirically from the 13 benchmark APIs and aligned with dependency classifications in prior work; added citations to RESTler, Morest, foREST, and Zhang et al.}
These five types --- Resource Lifecycle, Identifier Flow, Field-level Data Passing, Nested Resource Dependency, and State Propagation --- were derived from a pilot study on three APIs (\texttt{restcountries}, \texttt{features-service}, \texttt{market}; marked $^{\ddagger}$ in Table~\ref{TAB:APIs}) selected to cover stateless read-only, CRUD, and stateful workflow patterns, and they align with the producer-consumer relationships, resource-based ordering, and tree-structured endpoint hierarchies identified in prior REST API testing research~\cite{atlidakisRESTlerAutomaticIntelligenta,liuMorestModelbasedRESTful2022a,linFoRESTTreebasedBlackbox2023,zhangResourcebasedTestCase2019}.
The remaining 10 APIs serve as the held-out validation set, across which no dependency type outside these five was observed.
Cross-endpoint workflow constraints that fall outside these types are not modeled.

% The challenger review loop iterates until the reviewer emits no hints or the iteration count reaches a threshold $N$.
% The current experiments use $N = 1$; increasing $N$ trades token cost for oracle quality.
The current ablation study also fixes the model assignment (DeepSeek V4 Pro as primary, Qwen3.6-Plus as challenger); whether swapping the two assignments, using a different challenger model, or performing more than one challenger review pass per oracle generation yields further quality gains remains an open question for future work.

\subsection{Threats to Validity}
% \man{do we plan to open source the code and experiment data? The open-source oracle may be considered an additional contribution.}
% \dsd{Yes. Added contribution (6) in the Introduction: MASTOR source code, benchmark datasets, and the complete oracle suite for all 13 subject APIs will be released on GitHub. The oracle suite is highlighted as a reusable artifact and baseline for future research.}

\subsubsection{Internal Validity}

Internal validity concerns whether the observed mutation score differences are attributable to MASTOR's oracle generation rather than to confounding factors in the experimental setup.

\textbf{Implementation correctness.}
MASTOR coordinates multiple components across two phases.
Implementation defects in either phase may affect results.
To mitigate this, we use serializable intermediate representations (Source context and the oracle suite) that allow stepwise inspection at every phase boundary.
Oracle normalization enforces that every oracle references a valid evidence entry in the Source context, providing an automated consistency check.
We further manually verified oracle outputs for a subset of operations against the source code.
% {\wang{How was the manual inspection conducted?}}
% \dsd{Added description below.}
Specifically, three experienced software engineers independently inspected a randomly sampled 10\% of all endpoint operations (30 out of 296), checking whether each oracle's assertions were consistent with its evidence entries and the corresponding implementation source code.
No systematic inconsistencies or hallucinated assertions were identified, confirming that oracle normalization and evidence grounding effectively suppress unsupported outputs.

\textbf{LLM nondeterminism.}
Oracle generation involves LLM-based reasoning, and stochastic decoding may introduce run-to-run variability.
We set temperature = 0 for all inference calls where the model API supports it, and confirmed output stability by repeating the generation on selected operations and observing that the oracle sets remain structurally identical across runs.
Minor variation across model version updates may still occur, but is outside experimental control.

\textbf{Baseline fairness.}
The comparison with Direct Prompting and SATORI uses the same mutant set, test input suite, execution environment, and PostmanAssertify runtime for all three tools.
MASTOR, Direct Prompting, and SATORI all use the same underlying LLM (DeepSeek V4 Pro) or the oracle format documented by each tool's authors, with no additional optimization applied to any baseline.
This ensures that observed differences reflect oracle quality rather than execution or conversion artifacts.

\subsubsection{External Validity}

External validity concerns whether the results generalize beyond the specific subject APIs and evaluation context used in this study.

\textbf{Dataset representativeness.}
Subject APIs are drawn from the WFD~\cite{sahinWFCWFDWeb2025} and PRAB~\cite{decropPublicBenchmarkREST2025a} benchmarks, which are established community benchmarks used across more than 60 RESTful API testing studies.
The 13 selected APIs were chosen by frequency of citation across those studies, covering a range of application domains, database backends, and authentication configurations.
However, all subjects implement the Spring Boot layered architecture.
Generalization to industrial-scale microservice systems, polyglot stacks, or non-Java frameworks should be interpreted cautiously.
Evaluating MASTOR on a broader range of architectures is a direction for future work.

\textbf{Mutation operator realism.}
PITest mutation operators are general-purpose and may not cover fault classes specific to RESTful API business logic, such as incorrect status code mapping or missing field serialization.
Additionally, the API test input suite may fail to exercise some code paths, leaving a subset of reachable mutants without effective test coverage.
This could cause MS to underestimate the proportion of API behavior actually covered by the generated oracles.
To partially mitigate this, we exclude \textsc{no\_coverage} mutants from the MS denominator, restricting the score to mutants that the test inputs actually reach.

\subsubsection{Construct Validity}

Construct validity concerns whether the mutation score is an appropriate and accurate proxy for oracle effectiveness.

\textbf{Mutation score as an oracle quality proxy.}
Mutation score has been widely used to evaluate test oracle quality~\cite{jiaAnalysisMutationTesting2011}.
A higher MS indicates that more behavioral deviations introduced by mutations are detected, which reflects stronger assertion coverage.
However, MS measures killing rate on a specific mutant set, not the correctness of the asserted behavior in general.
Two oracles with identical MS may differ substantially in the classes of real faults they detect.
We treat MS as a comparative proxy rather than an absolute oracle quality measure.

\textbf{Equivalent mutants.}
\label{sec:threats}
Equivalent mutants are not explicitly identified or excluded.
PITest's \textsc{no\_coverage} mutants are excluded from the MS denominator, but \textsc{survived} mutants may include equivalent ones that no test input can semantically distinguish from the original.
This may cause MS to be slightly underestimated.
Explicit equivalent mutant detection is computationally hard and is omitted by most mutation testing studies~\cite{jiaAnalysisMutationTesting2011};
we treat this as a shared limitation of the evaluation methodology.

\textbf{Source analysis scope.}
The transitive import closure used in Source Analysis covers classes referenced through source-level Java imports.
Classes referenced only through reflection, dynamic proxies, or XML configuration descriptors are excluded from the source bundle.
If such classes encode endpoint-relevant constraints, those constraints are not visible to \texttt{SourceExtractionAgent}, which may cause oracle assertions to be incomplete.
This affects construct validity because the MS reflects oracle coverage only over constraints that are extractable by the current Source Analysis design.

% \section{Related Work
% \label{sec:9}}

% \subsection{AGORA+
% \label{sec:9.1}}

% \subsection{SATORI
% \label{sec:9.2}}

\section{Related Work
\label{sec:related}}

Research on test oracle generation for RESTful APIs has gained increasing attention in recent years~\cite{golmohammadiTestingRESTfulAPIs2023b}, as existing API testing tools often emphasize input generation while providing only limited support for semantic verification. 
Previous studies have investigated various strategies for automated oracle construction, predominantly leveraging API specifications.
% \xuan{not sure whether it is enough to only discuss two works in the related work section. I would recommend adding one more subsection about RESTful API testing.}
% \dsd{Added LlamaRestTest and AutoRestTest as separate subsections (each titled by the approach name, like AGORA+/SATORI) before the oracle-generation works, contrasting each with MASTOR.}
In this section, we first review two recent LLM-based approaches that focus on test input generation (LlamaRestTest and AutoRestTest), and then summarize two representative approaches that target test oracle generation (AGORA+ and SATORI).

\subsection{LlamaRestTest}

LlamaRestTest was proposed by Kim et al.~\cite{kimLlamaRestTestEffectiveREST2025a} as a small-language-model approach to RESTful API test generation.
The authors fine-tuned and quantized a Llama3-8B model on mined datasets of API example values and inter-parameter dependencies, and used the resulting models to produce realistic test inputs and to uncover inter-parameter dependencies by analyzing server responses.
Their evaluation on 12 real-world services showed that the fine-tuned small models matched or exceeded larger models in detecting parameter-dependency rules and generating valid inputs, while improving code coverage and the number of internal server errors found.

LlamaRestTest shares with MASTOR the use of language models, but it directs the model toward generating test inputs that drive deeper API exploration, whereas MASTOR directs the model toward generating test oracles that judge response correctness.
Because LlamaRestTest verifies behavior through interface-level signals such as HTTP status codes and internal server errors, it cannot detect semantic faults that return a well-formed response with incorrect content.
Our approach addresses this gap by grounding oracle generation in implementation source code, allowing it to assert behavior that input-generation tools leave unchecked.

\subsection{AutoRestTest}

AutoRestTest was proposed by Kim et al.~\cite{kimMultiAgentApproachREST2025a,stennettAutoRestTestToolAutomated2025b} as the first multi-agent approach for black-box RESTful API testing.
The approach combined multi-agent reinforcement learning with a semantic property dependency graph and an LLM, where four agents responsible for operations, dependencies, parameters, and values collaborated to explore the API.
The LLM generated domain-specific parameter values, the dependency graph reduced the search space for operation dependencies, and reinforcement learning optimized the joint exploration behavior.
The authors reported that AutoRestTest outperformed leading black-box testing tools on 12 real-world services in code coverage, operation coverage, and fault detection.

AutoRestTest shares with MASTOR both the use of LLMs and a multi-agent organization, which makes it the closest in design to our approach.
The two methods differ in their objective: AutoRestTest coordinates its agents to generate test inputs that maximize API exploration, whereas MASTOR coordinates its agents to generate test oracles from implementation source code.
Because AutoRestTest verifies behavior through interface-level signals such as code coverage and server errors, it cannot detect semantic faults whose responses are well-formed but incorrect in content.
Our approach complements input-driven exploration by providing the semantic oracles that determine whether an explored response is correct.

\subsection{AGORA+}

AGORA+ was proposed by Alonso et al.~\cite{alonsoTestOracleGeneration2025} as a dynamic test oracle inference method for RESTful APIs. 
The work addressed the observation that many API testing tools can generate valid requests but lack mechanisms to check deeper semantic properties of responses.
AGORA+ learned likely invariants from collected request and response traces and treated these invariants as candidate semantic constraints.
To support this process, the authors developed a transformation layer that converted API interactions into a format compatible with Daikon~\cite{ernstDaikonSystemDynamic2007}, a dynamic invariant detection system, extended for this purpose.
The inferred properties included value bounds, structural relations, and cross-field consistency constraints.
These constraints were subsequently translated into executable assertions through PostmanAssertify~\cite{alonsoAGORAAutomatedGeneration2023}, enabling their integration into automated testing workflows.
Experimental results on real-world APIs showed that AGORA+ could discover meaningful semantic properties and reveal defects that interface-level verification misses.

AGORA+ derived its constraints from observed executions; therefore, the coverage of inferred properties depended on the diversity of available traces.
Our approach, by contrast, reasons directly from implementation source code, allowing it to analyze control-flow conditions and state-dependent behaviors even when they are not exercised during testing.

\subsection{SATORI}
SATORI was introduced by Alonso et al.~\cite{alonsoSATORIStaticTest2025a} as a static oracle generation approach for RESTful APIs. 
Unlike dynamic invariant mining techniques, SATORI operated on OpenAPI specifications and leveraged LLMs to infer expected behavior by analyzing response schemas, field names, types, and descriptive metadata.
The approach generated structured assertions describing the semantic constraints on response properties, which were then translated into executable assertions with PostmanAssertify~\cite{alonsoAGORAAutomatedGeneration2023}.
Experimental results showed that SATORI could automatically produce a large number of valid semantic oracles per API operation and achieve competitive accuracy in detecting faults.
The authors further observed that static specification-based inference and dynamic invariant mining exhibit complementary strengths in practice.

Because SATORI derived oracles primarily from API specifications, it depended on the completeness and accuracy of documented behavior.
Semantic constraints encoded in implementation logic, such as control-flow branching and state-dependent behavior, may not be fully reflected in specifications.
Our approach addresses this limitation by analyzing implementation source code to extract structured semantic evidence for oracle generation.

\section{Conclusions and Future Work
\label{sec:conclusion}}

% {\wang{The conclusion description seems a bit verbose.}}
% \dsd{Revised and simplified this section.}
This paper presented MASTOR, a multi-agent approach to generating semantic test oracles for RESTful APIs grounded in implementation source code analysis.
By decomposing oracle generation into specialized agents that extract per-operation source evidence and synthesize assertions across both single-operation and multi-operation dimensions, MASTOR produces oracle suites that go beyond what API specifications or execution traces can provide.

The evaluation of 13 open-source RESTful API projects (296 endpoint operations, 251,303 lines of code) produces the following findings.
MASTOR generates 10,022 oracles and achieves an average mutation score of 75.4\% (range: 69.0\%--95.9\%) across all 13 APIs (RQ1).
On a comparable status-and-field oracle subset, MASTOR outperforms Direct Prompting by 30.1 pp (69.9\% vs.\ 39.8\%) and SATORI by 49.4 pp (69.9\% vs.\ 20.5\%), confirming the advantage of source-code-grounded multi-agent orchestration over unstructured prompting and specification-only inference (RQ2).
Ablation results show that Multi-Op Oracle Generation and ChallengerAgent each contribute independently, with their combined removal degrading average MS by 11.2 pp (RQ3).
The median inference cost is \$0.56 per API, with cost scaling predictably with endpoint count (RQ4).

Several directions extend the current work.
First, Source context extraction currently relies on synchronous, import-level static traversal; extending it to support reactive and event-driven frameworks, and to reach code accessed via Java reflection or runtime-generated proxies, would broaden its applicability.
Second, multi-operation oracle generation covers five static association types; incorporating dynamic execution traces and explicit state-transition modeling would surface additional cross-operation dependencies and enable workflow-level constraints spanning entire user journeys.
Third, richer static analysis that captures inter-field constraints and conditional serialization logic would enable more precise field-level assertions than the current evidence model supports.
Fourth, for large-scale systems, incremental re-extraction strategies that update only changed operations would reduce CI/CD re-analysis cost; additionally, investigating the effect of increasing the number of challenger review passes and varying the challenger model warrants empirical study.

\section*{Acknowledgment}
This work was partly supported by the Science and Technology Development Fund of Macau, Macau SAR (Grant Nos. 0069/2025/RIB2 and 0021/2023/RIA1) and Macau University of Science and Technology Faculty Research Grants (FRG) (Grant No. FRG-26-030-FIE).

\bibliographystyle{ACM-Reference-Format}
\bibliography{reference}

@article{jiaAnalysisMutationTesting2011,
  author    = {Jia, Yue and Harman, Mark},
  title     = {An Analysis and Survey of the Development of Mutation Testing},
  journal   = {IEEE Transactions on Software Engineering},
  volume    = {37},
  number    = {5},
  pages     = {649--678},
  year      = {2011}
}

@inproceedings{godefroidIntelligentRESTAPI2020a,
  title = {Intelligent {{REST API}} Data Fuzzing},
  booktitle = {Proceedings of the 28th {{ACM Joint Meeting}} on {{European Software Engineering Conference}} and {{Symposium}} on the {{Foundations}} of {{Software Engineering}}},
  author = {Godefroid, Patrice and Huang, Bo-Yuan and Polishchuk, Marina},
  year = {2020},
  pages = {725--736}
}

@article{diasFuzzTheRESTIntelligentAutomated2024,
  title = {{{FuzzTheREST}}: {{An Intelligent Automated Black-box RESTful API Fuzzer}}},
  author = {Dias, Tiago and Maia, Eva and Pra{\c c}a, Isabel},
  journal = {arXiv preprint},
  volume = {2407.14361},
  year = {2024}
}

@article{zhengRESTLessEnhancingStateoftheArt2024,
  title = {{{RESTLess}}: {{Enhancing State-of-the-Art REST API Fuzzing With LLMs}} in {{Cloud Service Computing}}},
  author = {Zheng, Tao and Shao, Jiang and Dai, Jinqiao and Jiang, Shuyu and Chen, Xingshu and Shen, Changxiang},
  year = {2024},
  journal = {IEEE Transactions on Services Computing},
  volume = {17},
  number = {6},
  pages = {4225--4238}
}

@inproceedings{rooijakkersWuppieFuzzCoverageGuidedStateful2026,
  title = {{{WuppieFuzz}}: {{Coverage-Guided}}, {{Stateful REST API Fuzzing}}},
  booktitle = {Proceedings of the 12th {{International Conference}} on {{Information Systems Security}} and {{Privacy}}},
  author = {Rooijakkers, Thomas and Nijsten, Anne and Daniele, Cristian and Weitenberg, Erieke and Groenewegen, Ringo and Melissen, Arthur},
  year = {2026},
  pages = {221--231}
}

@article{zhangOpenProblemsFuzzing2023a,
  title = {Open {{Problems}} in {{Fuzzing RESTful APIs}}: {{A Comparison}} of {{Tools}}},
  author = {Zhang, Man and Arcuri, Andrea},
  year = {2023},
  journal = {ACM Transactions on Software Engineering and Methodology},
  volume = {32},
  number = {6},
  pages = {144:1--144:45}
}

@inproceedings{fernandesCaseStudyApplying2025,
  title = {Case {{Study}}: {{Applying Fuzzing}} to {{REST APIs}} in a {{Large-Scale Industrial Setting}}},
  booktitle = {Proceedings of the {{IEEE}} 36th {{International Symposium}} on {{Software Reliability Engineering Workshops}} ({{ISSREW}})},
  author = {Fernandes, Leo and Rocha, Jo{\~a}o Vitor Souza and Wiese, Igor and Correia, Jo{\~a}o Lucas and Fraga, Ismael Trindade and Pinto, Renato Torres and Barboza, Erick and Fonseca, Baldoino and Ribeiro, M{\'a}rcio},
  year = {2025},
  pages = {45--50}
}

@article{arcuriAdvancedWhiteBoxHeuristics2024,
  title = {Advanced White-Box Heuristics for Search-Based Fuzzing of {REST APIs}},
  author = {Arcuri, Andrea and Zhang, Man and Galeotti, Juan},
  year = {2024},
  journal = {ACM Transactions on Software Engineering and Methodology},
  volume = {33},
  number = {6},
  pages = {142:1--142:36}
}

@inproceedings{golmohammadiEnhancingWhiteBoxSearchBased2023,
  title = {Enhancing {{White-Box Search-Based Testing}} of {{RESTful APIs}}},
  booktitle = {Proceedings of the {{IEEE}} 34th {{International Symposium}} on Software Reliability Engineering Workshops},
  author = {Golmohammadi, Amid},
  year = {2023},
  pages = {9--12}
}

@inproceedings{stallenbergImprovingTestCase2021a,
  title = {Improving {{Test Case Generation}} for {{REST APIs Through Hierarchical Clustering}}},
  booktitle = {Proceedings of the 36th {{IEEE}}/{{ACM International Conference}} on Automated Software Engineering},
  author = {Stallenberg, Dimitri and Olsthoorn, Mitchell and Panichella, Annibale},
  year = {2021},
  pages = {117--128}
}

@article{ghianniSearchBasedFuzzingRESTful2025,
  title = {Search-{{Based Fuzzing For RESTful APIs That Use MongoDB}}},
  author = {Ghianni, Hernan and Zhang, Man and Galeotti, Juan P. and Arcuri, Andrea},
  journal = {arXiv preprint},
  volume = {2507.20848},
  year = {2025}
}

@article{kimMultiAgentApproachREST2025a,
  title = {A {{Multi-Agent Approach}} for {{REST API Testing}} with {{Semantic Graphs}} and {{LLM-Driven Inputs}}},
  author = {Kim, Myeongsoo and Stennett, Tyler and Sinha, Saurabh and Orso, Alessandro},
  journal = {arXiv preprint},
  volume = {2411.07098},
  year = {2025}
}

@inproceedings{stennettAutoRestTestToolAutomated2025b,
  title = {{{AutoRestTest}}: {{A Tool}} for {{Automated REST API Testing Using LLMs}} and {{MARL}}},
  booktitle = {Proceedings of the {{IEEE}}/{{ACM}} 47th {{International Conference}} on {{Software Engineering}}: Companion Proceedings},
  author = {Stennett, Tyler and Kim, Myeongsoo and Sinha, Saurabh and Orso, Alessandro},
  year = {2025},
  pages = {21--24}
}

@article{hanMASTESTLLMBasedMultiAgent2025,
  title = {{{MASTEST}}: {{A LLM-Based Multi-Agent System For RESTful API Tests}}},
  author = {Han, Xiaoke and Zhu, Hong},
  journal = {arXiv preprint},
  volume = {2511.18038},
  year = {2025}
}

@article{kimLlamaRestTestEffectiveREST2025a,
  title = {{{LlamaRestTest}}: {{Effective REST API Testing}} with {{Small Language Models}}},
  author = {Kim, Myeongsoo and Sinha, Saurabh and Orso, Alessandro},
  year = {2025},
  journal = {Proceedings of the ACM on Software Engineering},
  volume = {2},
  number = {FSE},
  pages = {465--488}
}

@article{barradasCombiningTSLLLM2025,
  title = {Combining {{TSL}} and {{LLM}} to {{Automate REST API Testing}}: {{A Comparative Study}}},
  author = {Barradas, Thiago and Paes, Aline and Neves, V{\^a}nia de Oliveira},
  journal = {arXiv preprint},
  volume = {2509.05540},
  year = {2025}
}

@article{nooyensTestAmplificationREST2025,
  title = {Test {{Amplification}} for {{REST APIs}} via {{Single}} and {{Multi-Agent LLM Systems}}},
  author = {Nooyens, Robbe and Bardakci, Tolgahan and Beyazit, Mutlu and Demeyer, Serge},
  journal = {arXiv preprint},
  volume = {2504.08113},
  year = {2025}
}

@article{besjesAgenticLLMsREST2025,
  title = {Agentic {{LLMs}} for {{REST API Test Amplification}}: {{A Comparative Study Across Cloud Applications}}},
  author = {Besjes, Jarne and Nooyens, Robbe and Bardakci, Tolgahan and Beyazit, Mutlu and Demeyer, Serge},
  journal = {arXiv preprint},
  volume = {2510.27417},
  year = {2025}
}

@article{golmohammadiTestingRESTfulAPIs2023b,
  title = {Testing {{RESTful APIs}}: {{A Survey}}},
  author = {Golmohammadi, Amid and Zhang, Man and Arcuri, Andrea},
  year = {2023},
  journal = {ACM Transactions on Software Engineering and Methodology},
  volume = {33},
  number = {1},
  pages = {27:1--27:41}
}

@inproceedings{garciaAdvancesWebAPI2023,
  title = {Advances in {{Web API}} Testing: {{A Systematic Mapping Study}}},
  booktitle = {Proceedings of the 2023 {{Mexican International Conference}} on {{Computer Science}}},
  author = {Garc{\'i}a, Josue Capistran and Hern{\'a}ndez, Jorge Octavio Ochar{\'a}n and Arriaga, Juan Carlos Per{\'e}z and Ria{\~n}o, Hector Javier Lim{\'o}n},
  year = {2023},
  pages = {1--8}
}

@inproceedings{atlidakisRESTlerAutomaticIntelligenta,
  title = {RESTler: Stateful REST API Fuzzing},
  booktitle = {Proceedings of the 41st {{International Conference}} on {{Software Engineering}}},
  author = {Atlidakis, Vaggelis and Godefroid, Patrice and Polishchuk, Marina},
  year = {2019},
  pages = {748--758}
}

@article{zhangAdaptiveHypermutationSearchBased2021,
  title = {Adaptive {{Hypermutation}} for {{Search-Based System Test Generation}}: {{A Study}} on {{REST APIs}} with {{EvoMaster}}},
  author = {Zhang, Man and Arcuri, Andrea},
  year = {2022},
  journal = {ACM Transactions on Software Engineering and Methodology},
  volume = {31},
  number = {1},
  pages = {1--52}
}

@inproceedings{kimLeveragingLargeLanguage2024a,
  title = {Leveraging {{Large Language Models}} to {{Improve REST API Testing}}},
  booktitle = {Proceedings of the 2024 {{ACM}}/{{IEEE}} 44th {{International Conference}} on {{Software Engineering}}: {{New Ideas}} and {{Emerging Results}}},
  author = {Kim, Myeongsoo and Stennett, Tyler and Shah, Dhruv and Sinha, Saurabh and Orso, Alessandro},
  year = {2024},
  pages = {37--41}
}

@inproceedings{alonsoSATORIStaticTest2025a,
  title = {{SATORI}: Static Test Oracle Generation for {REST APIs}},
  booktitle = {Proceedings of the 40th {IEEE}/ACM International Conference on Automated Software Engineering},
  author = {Alonso, Juan C. and Martin-Lopez, Alberto and Segura, Sergio and Bavota, Gabriele and Ruiz-Cortés, Antonio},
  year = {2025},
  pages = {1364--1376}
}

@inproceedings{liuMorestModelbasedRESTful2022a,
  title = {Morest: Model-Based {{RESTful API}} Testing with Execution Feedback},
  booktitle = {Proceedings of the 44th {{International Conference}} on {{Software Engineering}}},
  author = {Liu, Yi and Li, Yuekang and Deng, Gelei and Liu, Yang and Wan, Ruiyuan and Wu, Runchao and Ji, Dandan and Xu, Shiheng and Bao, Minli},
  year = {2022},
  pages = {1406--1417}
}

@inproceedings{alonsoAGORAAutomatedGeneration2023,
  title = {{{AGORA}}: {{Automated Generation}} of {{Test Oracles}} for {{REST APIs}}},
  booktitle = {Proceedings of the 32nd {{ACM SIGSOFT International Symposium}} on {{Software Testing}} and {{Analysis}}},
  author = {Alonso, Juan C. and Segura, Sergio and {Ruiz-Cort{\'e}s}, Antonio},
  year = {2023},
  pages = {1542--1553}
}

@article{alonsoTestOracleGeneration2025,
  title = {Test {{Oracle Generation}} for {{REST APIs}}},
  author = {Alonso, Juan C. and Ernst, Michael D. and Segura, Sergio and {Ruiz-Cort{\'e}s}, Antonio},
  year = {2025},
  journal = {ACM Transactions on Software Engineering and Methodology},
  volume = {35},
  number = {1},
  pages = {19:1--19:37}
}

@article{dengLRASGenLLMbasedRESTful2026,
  title = {{{LRASGen}}: {{LLM-based RESTful API Specification Generation}}},
  author = {Deng, Sida and Huang, Rubing and Zhang, Man and Cui, Chenhui and Towey, Dave and Wang, Rongcun},
  year = {2026},
  journal = {ACM Transactions on Software Engineering and Methodology},
  doi = {10.1145/3810241}
}

@inproceedings{huangGeneratingRESTAPI2024,
  title = {Generating {{REST API}} Specifications through Static Analysis},
  booktitle = {Proceedings of the {{IEEE}}/ACM 46th International Conference on Software Engineering},
  author = {Huang, Ruikai and Motwani, Manish and Martinez, Idel and Orso, Alessandro},
  year = {2024},
  pages = {1--13},
  doi = {10.1145/3597503.3639137}
}

@misc{lercherGeneratingAccurateOpenAPI2024a,
  title = {Generating Accurate {{OpenAPI}} Descriptions from {{Java}} Source Code},
  author = {Lercher, Alexander and Macho, Christian and Bauer, Clemens and Pinzger, Martin},
  year = {2024},
  publisher = {arXiv},
  doi = {10.48550/arXiv.2410.23873}
}

@misc{openapispecificationOpenAPISpecification,
  title = {{{OpenAPI Specification}} v3.1.1},
  author = {{OpenAPI Initiative}},
  year = {2024},
  howpublished = {\url{https://spec.openapis.org/oas/latest.html}}
}

@misc{deepseekDeepSeek,
  title = {Models \& {{Pricing}} | {{DeepSeek API Docs}}},
  author = {{DeepSeek}},
  year = {2025},
  howpublished = {\url{https://api-docs.deepseek.com/quick_start/pricing/}}
}

@misc{qwen,
  title = {Qwen3.6-Plus},
  author = {{Aliyun}},
  year = {2026},
  howpublished = {\url{https://www.aliyun.com/product/tongyi}}
}

@phdthesis{fieldingArchitecturalStylesDesign2000,
  title = {Architectural Styles and the Design of Network-Based Software Architectures},
  author = {Fielding, Roy Thomas},
  year = {2000},
  school = {University of California, Irvine}
}

@article{barrOracleProblemSoftware2015,
  title = {The Oracle Problem in Software Testing: A Survey},
  author = {Barr, Earl T. and Harman, Mark and McMinn, Phil and Shahbaz, Muzammil and Yoo, Shin},
  year = {2015},
  journal = {IEEE Transactions on Software Engineering},
  volume = {41},
  number = {5},
  pages = {507--525}
}

@article{ernstDaikonSystemDynamic2007,
  title = {The {{Daikon}} System for Dynamic Detection of Likely Invariants},
  author = {Ernst, Michael D. and Perkins, Jeff H. and Guo, Philip J. and McCamant, Stephen and Pacheco, Carlos and Tschantz, Matthew S. and Xiao, Chen},
  year = {2007},
  journal = {Science of Computer Programming},
  volume = {69},
  number = {1},
  pages = {35--45}
}

@article{sahinWFCWFDWeb2025,
  title = {{{WFC}}/{{WFD}}: {{Web Fuzzing Commons}}, {{Dataset}} and {{Guidelines}} to {{Support Experimentation}} in {{REST API Fuzzing}}},
  author = {Sahin, Omur and Zhang, Man and Arcuri, Andrea},
  journal = {arXiv preprint},
  volume = {2509.01612},
  year = {2025}
}

@inproceedings{decropPublicBenchmarkREST2025a,
  title = {A Public Benchmark of {REST APIs}},
  booktitle = {Proceedings of the {IEEE}/{ACM} 22nd International Conference on Mining Software Repositories},
  author = {Decrop, Alix and Eraso, Sara and Devroey, Xavier and Perrouin, Gilles},
  year = {2025},
  pages = {421--433}
}

@misc{fayderRestcountries,
  title = {Restcountries},
  author = {{WebFuzzing}},
  year = {2025},
  howpublished = {\url{https://github.com/WebFuzzing/Dataset/tree/main/jdk_8_maven/cs/rest/original/restcountries}}
}

@misc{martinezFeaturesService,
  title = {Features-{{Service}}},
  author = {{WebFuzzing}},
  year = {2025},
  howpublished = {\url{https://github.com/WebFuzzing/Dataset/tree/main/jdk_8_maven/cs/rest/original/features-service}}
}

@misc{genomeNexusGitHub,
  title = {Genome-{{Nexus}}},
  author = {{WebFuzzing}},
  year = {2025},
  howpublished = {\url{https://github.com/WebFuzzing/Dataset/tree/main/jdk_8_maven/em/external/rest/genome-nexus}}
}

@misc{languagetoolGitHub,
  title = {{{LanguageTool}}},
  author = {{WebFuzzing}},
  year = {2025},
  howpublished = {\url{https://github.com/WebFuzzing/Dataset/tree/main/jdk_8_maven/cs/rest/original/languagetool}}
}

@misc{marketGitHub,
  title = {Market},
  author = {{WebFuzzing}},
  year = {2025},
  howpublished = {\url{https://github.com/WebFuzzing/Dataset/tree/main/jdk_11_maven/em/external/rest/market}}
}

@misc{proxyprintGitHub,
  title = {{{ProxyPrint}}},
  author = {{WebFuzzing}},
  year = {2025},
  howpublished = {\url{https://github.com/WebFuzzing/Dataset/tree/main/jdk_8_maven/cs/rest/original/proxyprint}}
}

@misc{catwatchGitHub,
  title = {Catwatch},
  author = {{WebFuzzing}},
  year = {2025},
  howpublished = {\url{https://github.com/WebFuzzing/Dataset/tree/main/jdk_8_maven/cs/rest/original/catwatch}}
}

@misc{personControllerGitHub,
  title = {Person-{{Controller}}},
  author = {{WebFuzzing}},
  year = {2025},
  howpublished = {\url{https://github.com/WebFuzzing/Dataset/tree/main/jdk_21_maven/cs/rest/person-controller}}
}

@misc{userManagementGitHub,
  title = {User-{{Management}}},
  author = {{WebFuzzing}},
  year = {2025},
  howpublished = {\url{https://github.com/WebFuzzing/Dataset/tree/main/jdk_8_maven/cs/rest/original/user-management}}
}

@misc{trackingSystemGitHub,
  title = {Tracking-{{System}}},
  author = {{WebFuzzing}},
  year = {2025},
  howpublished = {\url{https://github.com/WebFuzzing/Dataset/tree/main/jdk_11_maven/cs/rest/tracking-system}}
}

@misc{swaggerPetstoreGitHub,
  title = {Swagger-{{Petstore}}},
  author = {{WebFuzzing}},
  year = {2025},
  howpublished = {\url{https://github.com/WebFuzzing/Dataset/tree/main/jdk_8_maven/cs/rest/original/swagger-petstore}}
}

@inproceedings{colesPITPracticalMutation2016,
  title = {{{PIT}}: {{A Practical Mutation Testing Tool}} for {{Java}}},
  booktitle = {Proceedings of the 25th ACM SIGSOFT International Symposium on Software Testing and Analysis},
  author = {Coles, Henry and Laurent, Thomas and Henard, Christopher and Papadakis, Mike and Ventresque, Anthony},
  year = {2016},
  pages = {449--452}
}

@inproceedings{wuAutoGenEnablingNext2023,
  title = {{{AutoGen}}: {{Enabling Next-Gen LLM Applications}} via {{Multi-Agent Conversation}}},
  booktitle = {Proceedings of the 2nd {{Workshop}} on {{Agent Learning}} in {{Open-Endedness}} at {{NeurIPS}} 2023},
  author = {Wu, Qingyun and Bansal, Gagan and Zhang, Jieyu and Wu, Yiran and Li, Beibin and Zhu, Erkang and Jiang, Li and Zhang, Xiaoyun and Zhang, Shaokun and Liu, Jiale and Awadallah, Ahmed Hassan and White, Ryen W. and Burger, Doug and Wang, Chi},
  year = {2023}
}

@inproceedings{yaoReActSynergizingReasoning2023,
  title = {{{ReAct}}: {{Synergizing Reasoning}} and {{Acting}} in {{Language Models}}},
  booktitle = {Proceedings of the 11th {{International Conference}} on {{Learning Representations}}},
  author = {Yao, Shunyu and Zhao, Jeffrey and Yu, Dian and Du, Nan and Shafran, Izhak and Narasimhan, Karthik and Cao, Yuan},
  year = {2023}
}

@article{hayesrothBlackboardArchitectureControl1985,
  title = {A Blackboard Architecture for Control},
  author = {Hayes-Roth, Barbara},
  year = {1985},
  journal = {Artificial Intelligence},
  volume = {26},
  number = {3},
  pages = {251--321}
}

@misc{deepseekV4ProHF,
  title = {DeepSeek-V4-Pro},
  author = {{DeepSeek}},
  year = {2025},
  howpublished = {\url{https://huggingface.co/deepseek-ai/DeepSeek-V4-Pro}}
}

@misc{qwen36HF,
  title = {Qwen3.6-35B-A3B},
  author = {{Qwen Team}},
  year = {2025},
  howpublished = {\url{https://huggingface.co/Qwen/Qwen3.6-35B-A3B}}
}

@inproceedings{linFoRESTTreebasedBlackbox2023,
  author = {Lin, Jiaxian and Li, Tianyu and Chen, Yang and Wei, Guangsheng and Lin, Jiadong and Zhang, Sen and Xu, Hui},
  title = {foREST: A Tree-based Black-box Fuzzing Approach for RESTful APIs},
  booktitle = {In Proceedings of the 34th IEEE International Symposium on Software Reliability Engineering (ISSRE)},
  pages = {695--705},
  year = {2023}
}

@inproceedings{zhangResourcebasedTestCase2019,
  author = {Zhang, Man and Marculescu, Bogdan and Arcuri, Andrea},
  title = {Resource-based test case generation for RESTful web services},
  booktitle = {In Proceedings of the Genetic and Evolutionary Computation Conference (GECCO)},
  pages = {1426--1434},
  year = {2019}
}

@misc{mastorGitHub,
  author = {Deng, Sida},
  title = {MASTOR: Multi-Agent Semantic Test Oracle Generation for RESTful APIs},
  year = {2025},
  howpublished = {\url{https://github.com/Alysrazorr/MASTOR}}
}

\clearpage
\appendix

\setcounter{lstlisting}{0}
\renewcommand{\thelstlisting}{\thesection.\arabic{lstlisting}}

\setcounter{page}{1}
\pagenumbering{arabic}
\section*{Appendices}
\addcontentsline{toc}{section}{Appendices}

\section{Direct Prompting Prompt Template}
\label{appendix:dp-prompt}

Listing \ref{lst:dp-prompt} indicates the zero-shot prompt template used by the Direct Prompting (DP).
% The Direct Prompting (DP) baseline uses the following zero-shot prompt template.
The placeholders \texttt{\{oas\_ops\}} and \texttt{\{source\_code\}} are filled at runtime with the OAS-defined operation list and the concatenated source bundle for each subject API, respectively.

\begin{lstlisting}[basicstyle=\ttfamily\scriptsize, breaklines=true, frame=single, caption={Zero-shot prompt template used by the Direct Prompting (DP) baseline.}, label={lst:dp-prompt}]
Given the source code below, generate as many test oracles
as possible for every REST operation listed. Cover as many
execution paths as you can.

## OAS-defined operations (use these exact op_id values):
{oas_ops}

## Source code:
{source_code}

Return strictly the following JSON
(no markdown fences, pure JSON only):
{
  "operations": [
    {
      "op_id": "<HTTP_METHOD> <path>",
      "path_oracles": [
        {
          "path_index": 0,
          "description": "<brief description of this path>",
          "input": {
            "path": {},
            "query": null,
            "body": null,
            "headers": null
          },
          "assertions": [
            {"type": "status", "expected": 200},
            {"type": "field", "field_path": "result",
             "op": "equals", "expected": 3}
          ]
        }
      ]
    }
  ]
}

Assertion ops: equals | not_null | is_null | gte | lte |
               matches | type
Assertion types: status | field
\end{lstlisting}

\end{document}